\documentclass[11pt]{article}
\usepackage{amssymb,amsmath,amsfonts}
\usepackage{graphicx}
\usepackage{graphics}
\usepackage{eepic,epsfig}

\begin{document}

\title{Finite temperature fermionic condensate and energy-momentum tensor in cosmic string spacetime}
\author{W. Oliveira dos Santos$^{1}$%
\thanks{E-mail: wagner.physics@gmail.com} ,\thinspace\ E. R. Bezerra de Mello$^{1}$\thanks{	E-mail: emello@fisica.ufpb.br} \\
%EndAName\\
$^{1}$\textit{Departamento de F\'{\i}sica, Universidade Federal da Para\'{\i}ba}\\
\textit{58.059-970, Caixa Postal 5.008, Jo\~{a}o Pessoa, PB, Brazil}}
\maketitle

\begin{abstract}
Here we analyze the expectation value of the fermionic condensate and the energy-momentum tensor associated with a massive charged fermionic quantum  field with a nonzero chemical potential propagating in a magnetic-flux-carrying cosmic string in thermal equilibrium at finite temperature $T$. The expectation values of the fermionic condensate and the energy-momentum tensor are expressed as the sum of vacuum expectation values and the finite temperature contributions coming from the particles and antiparticles excitation.  The thermal expectations values of the fermionic condensate and the energy-momentum tensor are even periodic functions of the magnetic flux with period being the quantum flux, and also even functions of the chemical potential. Because the analyses of vacuum expectation of the fermionic condensate and energy-momentum tensor have been developed in literature, here we are mainly interested in the investigation of the thermal corrections. In this way we explicitly study how these observable behaves in the limits of low  and  high temperatures, and also for points near the string. Besides the analytical discussions, we included some graphs that exhibit the behavior of these observable for different values of the physical parameters of the model. 
\end{abstract}

\bigskip


\bigskip

\section{Introduction}
The primordial Universe has been underwent to different phases transitions. As consequence some topological defects have been formed after  Planck time \cite{Kibble,V-S}. These include domain walls, cosmic strings and monopoles. The cosmic strings are of special interest. They can be considered as lines of trapped energy density, analogous to defects such as vortex lines in superconductors. 

The gravitational effect produced by a straight, thin and static cosmic string is quite remarkable: a particle place at rest will be no attracted to it. There is no local gravity. In fact the spacetime around a cosmic string is locally but not global flat. The effect generated by a cosmic string is to produce a deficit planar angle in the two-dimensional sub-space orthogonal to it.  In quantum field theory the lack of global flatness induces non-zero vacuum expectation values (VEVs) for physical observable. Specifically the analysis of the vacuum polarization effects associated to scalar and fermionic fields in an idealized cosmic string spacetime have been developed in \cite{scalar}-\cite{Site11} and \cite{ferm}-\cite{Beze08f}, respectively. Moreover, the presence of  the string allows effects such as  particle-antiparticle pair production by a single photon and bremsstrahlung radiation from charged particles which are not possible in empty Minkowski space due to the conservation of linear momentum \cite{Skarzhinsky}.  

 The analysis of the induced current associated with a scalar massive charged quantum field in $(1+D)$-dimensional magnetic-flux-carrying cosmic string spacetime has been developed in \cite{Braganca_15}; moreover, the investigations of the vacuum expectation values of field squared and energy-momentum tensor (EMT) in this system have been carried out in \cite{Braganca_19}.  In addition, the analysis of induced fermionic current, $\langle j^\mu\rangle=\langle{\bar{\psi}}\gamma^\mu\psi\rangle$, in $(1+3)$-dimensional cosmic string spacetime has been investigated in \cite{Mello13a}, and the VEV of the EMT, $\langle T^\mu_\nu\rangle$, and the fermionic condensate (FC),  $\langle{\bar{\psi}}\psi\rangle $, in this model was developed in \cite{Mello14a}.

 The thermal corrections to the bosonic current densities, $\langle j_\mu\rangle $, in $(1+D)$-dimensional magnetic-flux-carrying cosmic string spacetime have been calculated in \cite{Mohammadi_16}. There it was  shown that the induced charge density is an odd function of chemical potential. So when this parameter is zero, the contribution of particle and antiparticle cancel each other. The chemical potential imbalances these contributions. As to the azimuthal current density it is a even function of this quantity. Moreover, the thermal effects on the bosonic field squared and EMT in the corresponding geometry, have been considered in \cite{Oliveira23}.  In the latter it was observed that these observables are even functions of the chemical potential and also on the magnetic fluxes with period equal to the quantum flux. We understand that this is an important topic, since for a cosmic string in the early stages of the cosmological expansion of the Universe,  the typical state of a quantum field is a state containing particles and anti-particles in thermal equilibrium at finite temperature $T$. As to thermal correction of the fermionic EMT in cosmic string spacetime, Linet, in \cite{Linet96}, has calculated the Euclidean thermal Green function for a spin$-1/2$ massless field in the absence of magnetic flux, assuming zero chemical potential. In the latter, the thermal Green function has been obtained by imposing anti-periodic condition on the Euclidean time with period $\beta=1/T$. So, here we propose to reanalyze this system in a more profound way. Considering a non-zero chemical potential  for the fermionic field and the electromagnetic interaction between this charged fields with the magnetic flux.  The evaluations of finite temperature corrections to the fermion condensate, charge and currents densities in a $(1+2)-$dimensional conical spacetime with a magnetic flux located at cone apex, have been developed in \cite{Braganca_16}. Considering a $(1+3)-$dimensional magnetic-flux-carrying cosmic string spacetime, the thermal corrections of the charge and current densities for a massive fermionic field with nonzero chemical potential, $\mu $, have been calculated in \cite{Azadeh_15}. 

 In this present paper we want to continue in this line of investigation analyzing the thermal corrections of the fermionic condensate, $\langle{\bar{\psi}}\psi\rangle _T$, and respective EMT, $\langle T^\mu_\nu\rangle_T$,   associated to a charged fermionic quantum  field with nonzero chemical potential, $\mu $, propagating in a $(1+3)-$dimensional magnetic-flux-carrying cosmic string spacetime. Specifically we want to investigate the influence of a non-zero temperature on both observables.

This paper is organized as follows: In section \ref{sec2} we introduce the model setup that we want to study, and present the normalized positive and negative energy solution of the Dirac equation. In section \ref{FCondensate} we calculate the thermal expectation value of the fermionic condensate (FC), $\langle{\bar{\psi}}\psi\rangle _T$. As we will see this quantity is even function of the magnetic flux and chemical potential, $\mu$. Considering $|\mu|<m$ we can develop a series expansion in the term dependent of temperature providing a more workable expression. The corresponding expression allows us to  present, analytically, the behavior of the FC for some asymptotic regimes of the parameters, such as low and high temperature and for points near the string. Also we analyze the FC for the case where $|\mu|>m$. In addition we present some graphs exhibiting the behavior of $\langle{\bar{\psi}}\psi\rangle _T$ as function of temperature and distance to the string's core. In section \ref{Energy-momentum} we calculate the thermal expectation value of all component of the EMT, $\langle T^\mu_\nu\rangle_T$, considering $|\mu|<m$.  In  section \ref{sec5}, we investigate the general properties obeyed by the thermal EMT, and analyses in detail various asymptotic regime of the thermal energy-density, including low and high temperature, points near the string for the case $|\mu|<m$. For the $|\mu|>m$, the energy-density is analyzed in this section too. Also we present some plots exhibiting the behavior of  $\langle T^0_0\rangle_T$ as function of temperature and the distance to the string's core.
 In section \ref{conc}, we provide our conclusions and more relevant remarks. We leave for the appendix \ref{energy_high_T}, specific details related with the calculations of the FC considering $|\mu|>m$, and the thermal energy density in the limit of high temperature, respectively. Throughout the paper we use natural units $G=\hbar =c=k_B=1$.

\section{The geometry and the fermionic modes}
\label{sec2}
In this section we present the background geometry that we want to consider in our analysis. It corresponds to the spacetime produced by an infinitely long straight cosmic string. Considering the string along the $z-$axis, and adopting 
cylindrical coordinates the line element associated with this geometry reads,
\begin{equation}
	ds^{2}=dt^{2}-dr^{2}-r^{2}d\phi ^{2}-dz{}^{2},  \label{ds21}
\end{equation}
with $0\leq\phi\leq 2\pi/q$, $r\geq 0$ and $(t, z)\in (-\infty, \ \infty)$.  The parameter $q\geq 1$ encodes the planar angle deficit, and is related to the mass  per unit length of the string, $\mu _{0}$, by $q^{-1}=1-4\mu_0$. 

In the presence of an external electromagnetic four-vector potential $A_{\mu }$, the quantum dynamic of a massive charged spin$-1/2$ field in curved spacetime is governed by the Dirac equation,
\begin{equation}
	i\gamma ^{\mu }{\mathcal{D}}_{\mu }\psi -m\psi =0\ ,\ {\mathcal{D}}_{\mu
	}=\partial _{\mu }+\Gamma _{\mu }+ieA_{\mu }  \  .  \label{Direq}
\end{equation}%
In the above equation $\gamma ^{\mu }$ represent the Dirac matrices in curved spacetime and $\Gamma _{\mu }$ the corresponding spin connection. Both matrices can be expressed in terms of flat spacetime Dirac matrices, $\gamma ^{(a)}$, by the relations below,
\begin{equation}
	\gamma ^{\mu }=e_{(a)}^{\mu }\gamma ^{(a)}\ ,\ \Gamma _{\mu }=\frac{1}{4}%
	\gamma ^{(a)}\gamma ^{(b)}e_{(a)}^{\nu }e_{(b)\nu ;\mu }\ .  \label{Gammamu}
\end{equation}%
In the above equations, $e_{(a)}^{\mu }$ corresponds to the tetrad basis satisfying the relation $e_{(a)}^{\mu }e_{(b)}^{\nu }\eta ^{ab}=g^{\mu \nu }$, with $\eta ^{ab}$ representing the Minkowski spacetime metric tensor.

In order to find these eigenfunctions, we will use the flat spacetime Dirac matrices below:
\begin{equation}
	\gamma ^{(0)}=\left(
	\begin{array}{cc}
		1 & 0 \\
		0 & -1%
	\end{array}%
	\right) ,\;\gamma ^{(a)}=\left(
	\begin{array}{cc}
		0 & \sigma _{a} \\
		-\sigma _{a} & 0%
	\end{array}%
	\right) ,\;a=1,2,3  \  ,  \label{gam0l}
\end{equation}%
with $\sigma _{1},\sigma _{2},\sigma _{3}$ being the Pauli matrices. We will take the tetrad basis in the form
\begin{equation}
	e_{(a)}^{\mu }=\left(
	\begin{array}{cccc}
		1 & 0 & 0 & 0 \\
		0 & \cos (q\phi ) & -\sin (q\phi )/r & 0 \\
		0 & \sin (q\phi ) & \cos (q\phi )/r & 0 \\
		0 & 0 & 0 & 1%
	\end{array}%
	\right) ,  \label{emua}
\end{equation}%
where the index $a$ identifies the rows of the matrix. With this choice we can write 
\begin{equation}
	\gamma ^{0}=\gamma ^{(0)},\;\gamma ^{l}=\left(
	\begin{array}{cc}
		0 & \sigma ^{l} \\
		-\sigma^{l} & 0
	\end{array}
	\right) , \ l=r, \ \phi, \ z  \label{gamcurved}
\end{equation}%
with the $2\times 2$ matrices given by,
\begin{equation}
	\sigma^r=\left(
	\begin{array}{cc}
		0 & e^{-iq\phi } \\
		e^{iq\phi } & 0%
	\end{array}%
	\right) ,\;\sigma^{\phi}=-\frac{i}{r}\left(
	\begin{array}{cc}
		0 & e^{-iq\phi } \\
		-e^{iq\phi } & 0%
	\end{array}%
	\right) ,\;\;\sigma^{z}=\left(
	\begin{array}{cc}
		1 & 0 \\
		0 & -1%
	\end{array}%
	\right) .  \label{betl}
\end{equation}%
As to the spin connection we have,
\begin{equation}
	\Gamma _{\mu }=\frac{1-q}{2}\gamma ^{(1)}\gamma ^{(2)}\delta _{\mu
	}^{2} \  . 
	\label{gammu}
\end{equation}%

The corresponding constant vector potentials are given below,
\begin{equation}
	A_{\mu}=-q\Phi/{2\pi}\delta^2_\mu \ . 
	\label{Vector}
\end{equation}
In the above expression, $\Phi$ correspond to the magnetic flux along the string's core. Though the magnetic field strength corresponding to (\ref{Vector}) vanishes, the nontrivial topology of the background geometry leads to Aharonov-Bohm-like effects on the VEVs of physical observables.

In order to evaluate the thermal corrections to the FC and the expectation value of the EMT, we need the complete set of normalized positive- and negative-energy fermionic mode-functions. A few years ago, in Ref. \cite{Mello13a}, we have presented this set, whose modes are specified by the quantum numbers $\chi=(\lambda ,k,j,s)$. These functions can be written in a compact form by,
\begin{equation}
	\psi _{\chi }^{(\pm )}(x)=C_{\chi }^{(\pm )}e^{\mp iEt+kz+iq(j-1/2)\phi
	}\left(
	\begin{array}{c}
		J_{\beta _{j}}(\lambda r) \\
		sJ_{\beta _{j}+\epsilon _{j}}(\lambda r)e^{iq\phi } \\
		\pm \frac{k-is\epsilon _{j}\lambda }{E\pm m}J_{\beta _{j}}(\lambda r) \\
		\mp s\frac{k-is\lambda \epsilon _{j}}{E\pm m}J_{\beta
			_{j}+\epsilon _{j}}(\lambda r)e^{iq\phi }
	\end{array}
	\right) \ ,  \label{psi+n}
\end{equation}
where $J_{\nu }(x)$ corresponds the Bessel function \cite{Abra}.\footnote{In \eqref{psi+n} we have adopted a regular solution at origin, $r=0$. Another possible approach is to impose the MIT bag boundary condition on a cylinder of radius $a$ coaxial to the string. In this case the general solution is a combination of Bessel and Neumann functions of the form $Z_{\beta_j}(\lambda r)=c_1J_{\beta_j}(\lambda r)+N_{\beta_j}(\lambda r)$ (e.g., see \cite{BezerradeMello:2010ii}). The coefficients $c_1$ and $c_2$ depend on the functions $J_{\beta_j}(\lambda r)$ and $N_{\beta_j}(\lambda r)$ on the cylinder. Then taking the limit $a\rightarrow0$, we arrive at $c_2/c_1\to0$ and, therefore, we re-obtain the Bessel function,  $J_{\beta_j}(\lambda r)$.} The fermionic wave-functions (\ref{psi+n}) are eigenfunctions of the projection of total angular momentum operator along the $z-$direction,
\begin{equation}
	\widehat{J}_{3}\psi _{\chi}^{(\pm )}=\left( -i\partial _{\phi }+i\frac{q}{%
		2}\gamma ^{(1)}\gamma ^{(2)}\right) \psi _{\chi }^{(\pm )}=qj\psi _{\chi
	}^{(\pm )}\ ,  \label{J3}
\end{equation}
with eigenvalues $j=\pm 1/2,\pm 3/2,\ldots $. Moreover the other quantum  numbers are defined by, $s=\pm 1$ and $\lambda \geq 0$. The index of the Bessel function in \eqref{psi+n} is given by
\begin{equation}
	\beta _{j}=q|j+\alpha |-\epsilon _{j}/2\ ,\;\alpha =eA_{\phi }/q=-\Phi/\Phi _{0},  \label{betaj}
\end{equation}%
with $\epsilon _{j}=\mathrm{sgn}(j+\alpha )$ and with $\Phi _{0}=2\pi /e$ being the flux quantum.
The energy associated with this mode is
\begin{equation}
E=	E_\chi=\sqrt{\lambda ^{2}+k^{2}+m^{2}} \  .  \label{E+}
	\end{equation}

The normlization constant $C_{\chi }^{(\pm )}$ in (\ref{psi+n}) is determined from the orthonormalization condition
\begin{equation}
	\int d^{3}x\sqrt{g^{(3)}}\ (\psi _{\chi }^{(r)})^{\dagger }\psi _{\chi^{\prime }}^{(r^{\prime })}=\delta _{\chi \chi ^{\prime }}\ \delta
	_{rr^{\prime }},\;r,r^{\prime }=+,-,  \label{normcond}
\end{equation}%
where $g^{(3)} $ is the determinant of the spatial metric tensor. The delta symbol on the right-hand side is understood as the Dirac delta function for continuous quantum numbers ($\lambda, k $) and the Kronecker delta for discrete ones ($j,s,r$). From (\ref{normcond}) one finds
\begin{equation}
	|C_{\chi }^{(\pm )}|^{2}=\frac{q\lambda (E_\chi\pm m)}{16\pi^2 E_\chi}\ .  \label{C_Nor}
\end{equation}%

Here, we are interested in the thermal effects on the FC and EMT induced by the presence of magnetic-flux-carrying cosmic string. So we assume  that the field is in thermal equilibrium at finite temperature $T$. We will adopt the same procedure as in \cite{Bell14T}. The standard form of the density matrix for the thermodynamical equilibrium distribution at temperature $T$ is
\begin{equation}
	\hat{\rho}=Z^{-1}e^{-\beta (\hat{H}-\mu ^{\prime }\hat{Q})},\;\beta =1/T,
	\label{rho}
\end{equation}%
being $\hat{H}$ the Hamilton operator, $\hat{Q}$ the conserved charge and $\mu ^{\prime }$ the corresponding chemical potential. The grand canonical partition function $Z$ is given by
\begin{equation}
	Z=\mathrm{tr}[e^{-\beta (\hat{H}-\mu ^{\prime }\hat{Q})}]  \  .  \label{PartFunc}
\end{equation}

Considering the complete set of normalized positive- and negative-energy solutions of (\ref{Direq}), specified by a set of quantum numbers $\chi $, $\{\psi _{\chi }^{(+)},\psi _{\chi }^{(-)}\}$, to evaluate the FC and EMT, we expand the field operator as
\begin{equation}
	\psi =\sum_{\chi }[\hat{a}_{\chi }\psi _{\chi }^{(+)}+\hat{b}_{\chi }^{+}\psi _{\chi }^{(-)}]\ ,  \label{psiexp}
\end{equation}
where we use the compact notation defined as
\begin{equation}
	\sum_{\chi }=\int_{0}^{\infty }d\lambda \ \int \ dk\sum_{s=\pm 1}\sum_{j=\pm 1/2,\cdots }\ .  \label{Sumsig}
\end{equation}
In \eqref{psiexp}, $\hat{a}_{\chi }$ and $\hat{b}_{\chi}^{+}$ correspond the annihilation and creation operators associated to particles and antiparticles respectively, and use the relations
\begin{eqnarray}
	\mathrm{tr}[\hat{\rho}\hat{a}_{\chi }^{+}\hat{a}_{\chi ^{\prime }}] &=&	\frac{\delta _{\chi \chi ^{\prime }}}{e^{\beta (E_{\chi}-\mu )}+1},  \notag \\
	\mathrm{tr}[\hat{\rho}\hat{b}_{\chi}^{+}\hat{b}_{\chi ^{\prime }}] &=&%
	\frac{\delta _{\chi \chi ^{\prime }}}{e^{\beta (E _{\chi}+\mu )}+1}  \  ,  \label{traa}
\end{eqnarray}%
where $\mu =e\mu ^{\prime }$.

\section{Fermionic condensate}
\label{FCondensate}

In this section we want to investigate the combined effects of finite temperature, conical topology and electromagnetic interactions on the FC, considering that the charged field is in thermal equilibrium at finite temperature $T$. This quantity is given by,
\begin{equation}
	\left\langle \bar{\psi}\psi \right\rangle =\mathrm{tr}[\widehat{\rho }\bar{%
		\psi}\psi ],  \label{FC}
\end{equation}%
where $\bar{\psi}=\psi ^{+}\gamma ^{0}$ is the Dirac conjugated spinor, $\hat{\rho}$ is the density matrix and $\langle ... \rangle$ means the ensemble average.

Substituting \eqref{psiexp} into the above thermal average and using \eqref{traa}, we obtain, 
\begin{eqnarray}
	\label{FC}
	\left\langle \bar{\psi}\psi \right\rangle=	\left\langle \bar{\psi}\psi \right\rangle^{(0)}+	\left\langle \bar{\psi}\psi \right\rangle^{+} +	\left\langle \bar{\psi}\psi \right\rangle^{-} \  ,  
\end{eqnarray}
where the first term on the right hand side of \eqref{FC} corresponds the VEV of the FC,
\begin{eqnarray}
	\left\langle \bar{\psi}\psi \right\rangle^{(0)}=\sum_\chi {\bar{\psi}}_\chi^{(-)}\psi_\chi^{(-)}  \  , 
\end{eqnarray}
and the two other terms represent the contribution comming from particles $(+)$ and anti-particles $(-)$:
\begin{eqnarray}
\left\langle \bar{\psi}\psi \right\rangle^{\pm}=\pm\sum_\chi\frac{{\bar{\psi}}_\chi^{(\pm)}\psi_\chi^{(\pm)} }{e^{\beta(E_\chi\mp\mu)}+1}   \  . 
\end{eqnarray}

Defining, 
\begin{eqnarray}
	\label{FC1}
\left\langle \bar{\psi}\psi \right\rangle_T =	\left\langle \bar{\psi}\psi \right\rangle^+ +	\left\langle \bar{\psi}\psi \right\rangle^-   \  ,
\end{eqnarray}
and substituting the positive- negative-energy fermionic modes,  given in \eqref{psi+n} into  the corresponding contribution, we find,
\begin{eqnarray}
	\label{FCthermal}
\left\langle \bar{\psi}\psi \right\rangle_T	=\frac{mq}{(2\pi)^2}\sum_j\int\ dk\int_0^\infty \frac{\lambda}{E_\chi} \left[J_{\beta_j}^2(\lambda r)+J_{\beta_j+\epsilon_{j}}^2(\lambda r)\right]\sum_{\delta=\pm 1}\frac1{e^{\beta(E_\chi-\delta \mu)}+1}  \  ,
\end{eqnarray}
where the component with $\delta=+1$ corresponds the contribution from the particles, and with $\delta=-1$, from the anti-particles, as exhibited in \eqref{FC1}. At this point we would like to say that the thermal FC is a even function of chemical potential, and for $\mu=0$, the contributions from particle and anti-particle coincide. Also from \eqref{FCthermal} follows that the FC is an even periodic function of the parameter $\alpha$. Presenting $\alpha$ as,
\begin{eqnarray}
	\label{alpha}
	\alpha=N+\alpha_0, \ {\rm with} \  |\alpha_0|<1/2  \  ,
\end{eqnarray}
being $N$ an integer number, the thermal FC is a function only  of $\alpha_0$, the fractional part of the ratio of the magnetic flux to the quantum one. 

\begin{figure}[!htb]
	\begin{center}
		\centering
		\includegraphics[scale=0.365]{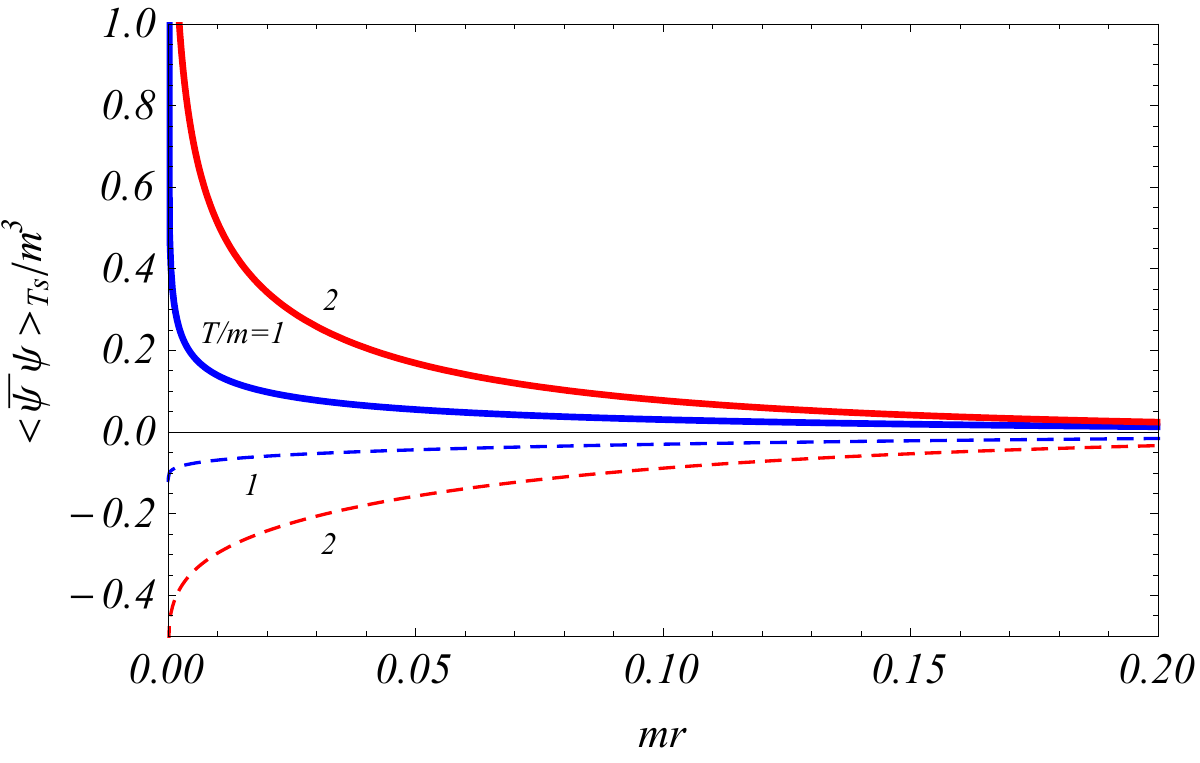}
		\quad
		\includegraphics[scale=0.35]{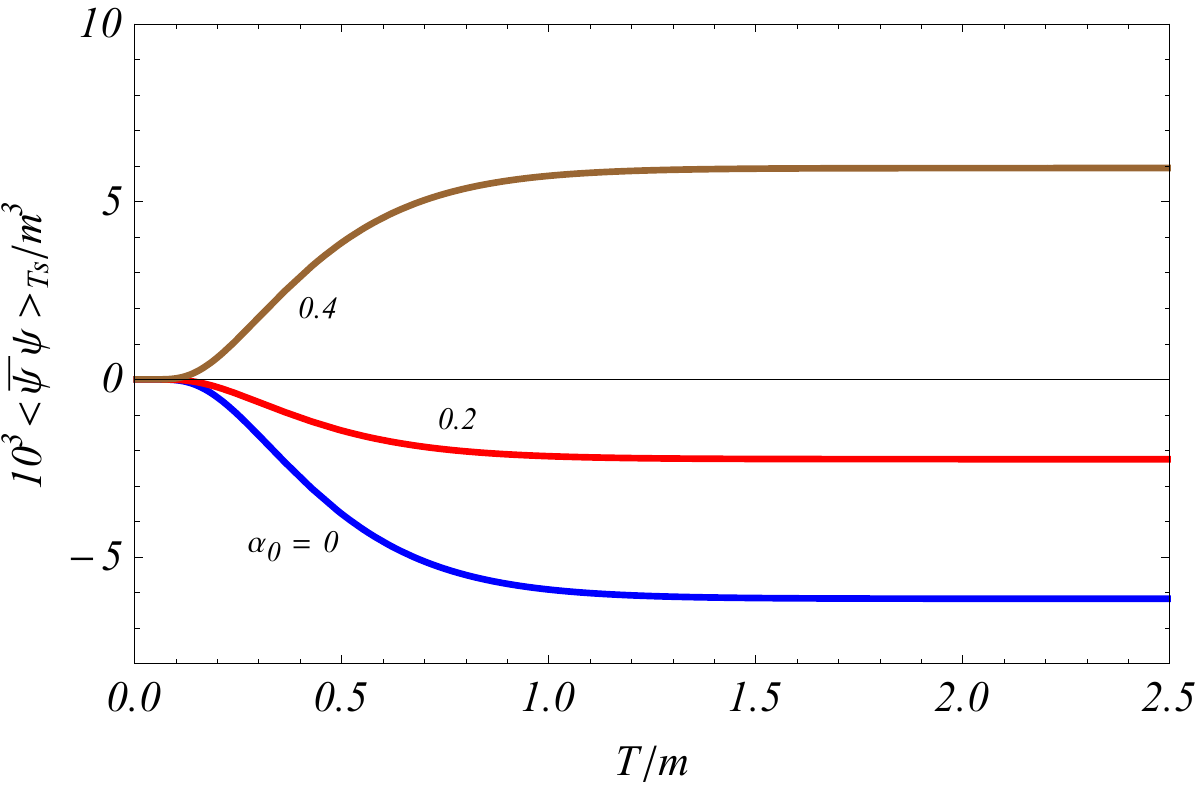}
		\caption{The FC induced by the cosmic string as a function of the product $mr$ (left panel) and the ratio $T/m$ (right panel). In both plots we have fixed the ratio $\mu/m=0.5$. Moreover, in the left panel plot we have fixed $\alpha_0=0.25$, while for the one in the right we have fixed $mr=1$ and $q=2.5$. In the left panel the full curves are given for $q=1.5$, while the dashed ones are given for $q=2.5$. The numbers near these curves represent different values of $T/m$, while numbers near the curves in the right panel represent different values of $\alpha_0$.}
		\label{fig1}
	\end{center}
\end{figure}

For further analysis  of the thermal FC, we first consider the case $|\mu|<m$. Accepting this condition we can use the expansion,
\begin{eqnarray}
	\label{expansion}
	\frac1{e^y+1}=-\sum_{n=1}^\infty(-1)^ne^{-ny} \  
\end{eqnarray}
in \eqref{FCthermal}, and obtain a more workable expression given below,\footnote{Due to the difficulty of finding an analytic expression for the FC departing from Eq. \eqref{FCthermal}, we had to adopt the series expansion in \eqref{expansion}, which is valid for $E_{\chi}-\delta\mu$ is positive. This surely happens for $m>|\mu|$. Moreover, our results are supported by the graphs displayed in Fig. \ref{fig1}, which were obtained for $n=20$.}
\begin{eqnarray}
	\label{FC_uncomp}
	\left\langle \bar{\psi}\psi \right\rangle_T&=&-\frac{mq}{2\pi^2}\sum_{n=1}^\infty\sum_{\delta=\pm1}(-1)^n e^{n\delta\beta\mu} \sum_j \int_0^\infty d\lambda{\lambda} \left[J_{\beta_j}^2(\lambda r)+J_{\beta_j+\epsilon_{j}}^2(\lambda r)\right]\nonumber\\
	&\times&	\int_0^\infty dk \frac{e^{-n\beta\sqrt{m^2+\lambda^2+k^2}}}{\sqrt{m^2+\lambda^2+k^2}} \  .
\end{eqnarray}

Using the identity,
\begin{eqnarray}
	\label{identity2}
	\frac{e^{-\sigma\sqrt{p^2+v^2}}}{\sqrt{p^2+v^2}}=\frac2{\sqrt{\pi}}\int_0^\infty ds e^{-(p^2+v^2)s^2-\sigma^2/(4s^2) }  \  ,
\end{eqnarray}
it is possible to integrate over $\lambda$ in \eqref{FC_uncomp},  by using the expression below \cite{Grad},
\begin{eqnarray}
	\label{Int_Bessel_J}
	\int_0^\infty d\lambda \lambda e^{-\lambda^2 s^2}J^2_\mu(\lambda r)=\frac{e^{-r^2/(2s^2)}}{2s^2}I_\mu\left(\frac{r^2}{2s^2}\right) \  , 
\end{eqnarray}
where $I_\mu(z)$ represents the modified Bessel function \cite{Abra}. The result is,
\begin{eqnarray}
\label{FC_uncomp_1}
\left\langle \bar{\psi}\psi \right\rangle_T&=&-\frac{mq}{4\pi^2}\sum_{n=1}^\infty\sum_{\delta=\pm1}(-1)^n e^{n\delta\beta\mu} \sum_j \int_0^\infty \frac{ds}{s^3}e^{-m^2s^2-n^2\beta^2/(4s^2)}\nonumber\\
&\times& e^{-r^2/(2s^2)}\left[I_{\beta_j}( r^2/(2s^2))+I_{\beta_j+\epsilon_{j}}(r^2/(2s^2))\right] \  .
\end{eqnarray}

Defining a new variable $x=\frac{r^2}{2s^2}$, the thermal FC above can be expressed by,
\begin{eqnarray}
\label{FC_uncomp_2}
\left\langle \bar{\psi}\psi \right\rangle_T&=&-\frac{mq}{(2\pi r)^2}\sum_{n=1}^\infty\sum_{\delta=\pm1}(-1)^n e^{n\delta\beta\mu} \int_0^\infty dx e^{-m^2r^2/(2x)-n^2\beta^2x/(2r^2)} \nonumber\\
&\times&e^{-x}{\cal{F}}(q,\alpha_0,x)  \  , 
\end{eqnarray}
where we defined 
\begin{eqnarray}
	\label{Sum_Bessel_I}
\sum_j\left[I_{\beta_j}(x)+I_{\beta_j+\epsilon_{j}}(x)\right]={\cal{F}}(q,\alpha_0,x)  \  .
\end{eqnarray}

In Ref. \cite{Beze10b} it was obtained an integral representation for the function ${\cal{F}}(q,\alpha_0,x)$ above. It reads,
\begin{eqnarray}
	\mathcal{F}(q,\alpha _{0},x) &=&\frac{4}{q}\left[\frac{e^{x}}2+\sum_{k=1}^{[q/2]}(-1)^{k}\cos(\pi k/q)\cos (2\pi k\alpha
	_{0})e^{x\cos (2\pi k/q)}\right.\nonumber\\
	&+&\left.\frac q\pi 
	\int_{0}^{\infty }dy\,\frac{h(q,\alpha _{0},y)\sinh y}{\cosh (2qy)-\cos(q\pi )}e^{-x\cosh (2y)} \right]\ ,  \label{Sum01}
\end{eqnarray}%
where $[q/2]$ represents the integer part of $q/2$. In the case $1\leq q<2$, the contribution with sum is absence. The function $h(q,\alpha_0,x)$ in the
integrand of (\ref{Sum01}) is given by the expression
\begin{eqnarray}
	\label{h_func}
	h(q,\alpha _{0},x) &=&\cos \left[ q\pi \left( 1/2+\alpha _{0}\right) \right]
	\sinh \left[ \left( 1-2\alpha _{0}\right) qx\right]  \notag \\
	&&+\cos \left[ q\pi \left( 1/2-\alpha _{0}\right) \right] \sinh \left[
	\left( 1+2\alpha _{0}\right) qx\right] \ .  \label{g0}
\end{eqnarray}
Moreover we can see that ${\cal{F}}(q,\alpha_0,x)$ is an even function of $\alpha _{0}$.

Note that the first term in  the right hand side of \eqref{Sum01} provides a contribution into \eqref{FC_uncomp_2} that is independent of $\alpha_0$ and $q$. This corresponds the thermal FC in Minkowski spacetime.

 Using the integral representation below for the modified Bessel function \cite{Abra},
\begin{eqnarray}
	\label{Rep_1}
K_{1+d}(2ab)=\frac12\left(\frac b a\right)^{d+1}\int_0^\infty dx \ x^d \ e^{-a^2/x-b^2 x}  \  ,
\end{eqnarray}
we obtain,
\begin{eqnarray}
	\label{FC_Min}
\left\langle \bar{\psi}\psi \right\rangle_T^{(M)}=-\frac{2m^3}{\pi^2}\sum_{n=1}^\infty(-1)^n\cosh(n\beta\mu)f_1(n\beta m) \  , 
\end{eqnarray}
with the notation,
\begin{eqnarray}
	\label{f-function}
f_\nu(x)=\frac{K_\nu(x)}{x^\nu} \   .
\end{eqnarray}

As to the contribution induced by the cosmic string, after many intermediate steps and using the integral representation \eqref{Rep_1}, we obtain
\begin{eqnarray}
	\label{FC_cs1}
	\left\langle \bar{\psi}\psi \right\rangle_{Ts}&=&-\frac{4m^3}{\pi^2}\sum_{n=1}^\infty(-1)^n\cosh(n\beta\mu) \left[\sum_{k=1}^{[q/2]}(-1)^{k}c_k\cos (2\pi k\alpha_{0})f_1\left(m u_{kn}\right)\right.\nonumber\\
	&+&\left.\frac q\pi \int_{0}^{\infty }dy\,\frac{h(q,\alpha _{0},y)\sinh y}{\cosh (2qy)-\cos(q\pi )}  f_1\left(m u_{yn}\right) \right]   \  ,
\end{eqnarray}
where we have defined the new variables
\begin{eqnarray}
	u_{kn}&=&\sqrt{4r^2s_k^2+(n\beta)^2} \nonumber\\
	u_{yn}&=&\sqrt{4r^2c_y^2+(n\beta)^2}  \  .
	\label{arguments}
\end{eqnarray}
and the notations
\begin{eqnarray}
	\label{index}
	s_k=\sin(\pi k/q) \ , \ c_y=\cosh(y) \ {\rm and} \ c_k=\cos(\pi k/q) \  .
\end{eqnarray}
In \eqref{FC_cs1} the contribution associated with particle corresponds to half of the component with $e^{n\mu\beta}$ and anti-particle the half of the component with $e^{-n\mu\beta}$.

Let us analyze the behavior of the thermal FC, $\left\langle \bar{\psi}\psi \right\rangle_{Ts}$, for some limiting cases. First we consider the region near the string. For $\sigma=(2|\alpha_0|-1)q+1<0$, the FC is finite on the string's core. In this case we can take $r=0$ directly in \eqref{FC_cs1}. The result is:
\begin{eqnarray}
\label{FC_cs2}
	\left\langle \bar{\psi}\psi \right\rangle_{Ts}=-\frac{4m^3}{\pi^2}\sum_{n=1}^\infty(-1)^n\cosh(n\beta\mu) f_1(mn\beta)g(q,\alpha_0) \  ,
\end{eqnarray}
with 
\begin{eqnarray}
	\label{g_function}
	g(q,\alpha_0)=\sum_{k=1}^{[q/2]}(-1)^{k}c_k\cos (2\pi k\alpha_{0})
	+\frac q\pi \int_{0}^{\infty }dy\,\frac{h(q,\alpha _{0},y)\sinh y}{\cosh (2qy)-\cos(q\pi )} \   .
\end{eqnarray}

In the case $\sigma>0$, the thermal FC diverges on the string. The divergence comes from the integral in \eqref{FC_cs1}. We can evaluate its behavior by analyzing the asymptotic expression for the integrand for large value of $y$. In this regime,
\begin{eqnarray}
\frac{h(q,\alpha_0,y)\sinh(y)}{\cosh(2qy)-\cos(q\pi)}\approx\frac{\cos[q\pi(1/2-|\alpha_0|)]}{2}  e^{\sigma y} \  .
\label{asymp_y}
\end{eqnarray}
Defining a new variable $x=mre^{y}$, the argument of the $f_1$ in \eqref{FC_cs1} becomes $\sqrt{(nm\beta)^2+x^2}$. Using the integral representation for the Macdonald function, \eqref{Rep_1}, for a general order of function $f_\nu$, we have:
\begin{eqnarray}
	\label{int_repr}
	f_\nu(\sqrt{(nm\beta)^2+x^2})= 2^{\nu-1}\int_0^\infty dz z^{\nu-1}e^{-1/(4z)-[(mn\beta)^2+x^2]z}   \  .
\end{eqnarray}
Substituting these results in the integral of \eqref{FC_cs1}, after a few intermediate steps more, we obtain,
\begin{eqnarray}
\label{FC_cs3}
\left\langle \bar{\psi}\psi \right\rangle_{Ts}&\approx&-\frac{qm^3}{\pi^3}\frac{2^{\sigma/2}}{(mr)^\sigma}\Gamma\left(\sigma/2\right)\cos((1/2-|\alpha_0|)q\pi) \sum_{n=1}^\infty(-1)^n \cosh(n\beta\mu)(nm\beta)^{\sigma/2-1}\nonumber\\
&\times&K_{\sigma/2-1}(nm\beta) \  .
\end{eqnarray}
Note that this divergence is associated with the presence of the string and its magnetic flux; for a fixed magnetic flux, $\alpha_{0}$, the FC can be divergent or finite according to the value of the parameter associated with the angular deficit, $q$. The plot in Fig.\ref{fig1} exemplify this. For a fixed $\alpha_{0}=0.25$, the FC induced by the string is divergent for $q=1.5$ ($\sigma>0$), while it is finite for $q=2.5$ ($\sigma<0$). It allow us to conclude that as the angular deficit increases, the properties of the vacuum tends to be more stable around the string's core.

Now let us analyze the thermal FC in the limit of low and high temperature. At low temperature regime, $T<m,r^{-1}$, and considering  $\sigma<0$, we express $\beta=1/T$  in the arguments of the functions $f_1(m\sqrt{n^2\beta^2+4r^2 s_k^2})$ and $f_1(m\sqrt{n^2\beta^2+4r^2 c_y^2})$ in \eqref{FC_cs1}. Factorizing $1/T$ and taking $Tr=0$, due the exponential decay of the Macdonald  function for large argument, the leading term contribution comes from $n=1$. After some intermediate steps we get,
\begin{eqnarray}
\label{FC_cs4}
\left\langle \bar{\psi}\psi \right\rangle_{Ts}\approx 8\left(\frac{mT}{2\pi}\right)^{3/2}e^{-(m-|\mu|)/T} g(q,\alpha_0) \  , 
\end{eqnarray}
being $g(q,\alpha_0)$ given by \eqref{g_function}. For $\sigma>0$ we cannot simply take $Tr=0$, because the integral term in \eqref{FC_cs1} diverges; however we can evaluate the behavior of the thermal FC considering first the  asymptotic expression for the integrand for large values of $y$. Factorizing $1/T$ in the argument of $f_1(m\sqrt{n^2\beta^2+4r^2 c_y^2})$ in \eqref{FC_cs1}, and defining a new variable $x=rTe^y$, we obtain after some algebraic manipulations, the following expression:\footnote{In fact the algebraic manipulations adopted here are similar to the ones in the analysis of FC for points near the string.}
\begin{eqnarray}
	\label{FC_cs5}
	\left\langle \bar{\psi}\psi \right\rangle_{Ts}\approx\frac{qm^3}{\pi^{5/2}}\frac{2^{(\sigma-3)/2}}{(rT)^\sigma} \Gamma(\sigma/2)\cos((1/2-|\alpha_0|)q\pi)\left(\frac Tm\right)^{(\sigma+3)/2} e^{-(m-|\mu|)/T} \  .  
\end{eqnarray} 

Now let us analyses the FC in the regime of high temperature. The main contributions come from large values of $n$ in \eqref{FC_cs1}. However, this representation is not convenient to develop the sum. In order to provide a more convenient expression to evaluate this behavior, we will modify the form of the summation by using the identity below \cite{Bellucci:2009jr, Cruz:2018bqt}:
\begin{eqnarray}
	\label{Resum}
	\sum_{n=1}^{\infty}\cos(\alpha n)f_{\nu}\left(m\sqrt{b^2+\beta^2n^2}\right)=-\frac{1}{2} f_{\nu}\left(mb\right)+\frac{1}{2}\frac{\sqrt{2\pi}}{\beta m^{2\nu}}\sum_{n=-\infty}^{\infty}w_{n}^{2\nu-1}f_{\nu-1/2}(b w_n) \ ,
\end{eqnarray}
where $b>0$ and $w_{n}=\sqrt{(2\pi n+\alpha)^2/\beta^2+m^2}$. For our case, we have $\alpha=\pi+ i\beta{\mu}$,  $b= \{ 2r\sin(\pi k/q), \ 2r\cosh(y)\}$ and $\nu=1$. In this way, we get $w_n=\sqrt{\left(\frac{2\pi n}{\beta}+\frac{\pi}{\beta}+i\mu\right)^2+m^2}$. So, in the high temperature regime, the dominant contributions in the second term on the right hand side of \eqref{Resum} come from $n=0$ and $n=-1$. So we can write 
\begin{eqnarray}
\sum_{n=1}^{\infty}\cos(\alpha n)f_{\nu}\left(m\sqrt{b^2+\beta^2n^2}\right)\approx\frac{T\pi}{m^2}\frac1b{\cal{R}}\left\{ e^{-b\sqrt{(\pi T+i\mu)^2+m^2}} \right\}\  , 
\end{eqnarray}
where ${\cal}{R}\{f(z)\}$ represents the real part of the function $f(z)$. So, we can rewrite the FC in the regime of high temperature by,
\begin{eqnarray}
	\label{high_Temp}
	\left\langle \bar{\psi}\psi \right\rangle_{Ts}&\approx&-\frac{2mT}{\pi r}\left[\sum_{k=1}^{[q/2]}(-1)^{k}\cot(\pi k/q)\cos (2\pi k\alpha_{0})e^{-2r\sin(\pi k/q)\pi T} \cos\left(2r\sin(\pi k/q)\mu\right)\right.\nonumber\\
	&+&\left.\frac{2 q}\pi \int_{0}^{\infty }dy\,\frac{h(q,\alpha _{0},y)\tanh(y)}{\cosh (2qy)-\cos(q\pi )} e^{-2r\cosh(y)\pi T}\cos\left(2r\cosh(y)\mu\right)\right]	\  .
\end{eqnarray}

Now let us investigate the behavior of  the thermal FC in the regime of large distance from the string, i.e., for $r\gg T^{-1}, \ m^{-1}$. The procedure to develop this analysis is similar to one used for high temperature regime. Adopting the identity \eqref{Resum}, we can see that the dominant contributions come from terms with $n=0$ and $n=-1$. So, we can write:
\begin{eqnarray}
	\label{high_dist}
	\left\langle \bar{\psi}\psi \right\rangle_{Ts}&\approx&-\frac{2mT}{\pi r}\left[\sum_{k=1}^{[q/2]}(-1)^{k}\cot(\pi k/q)\cos (2\pi k\alpha_{0}) {\cal{R}}\left\{ e^{-2r\sin(\pi k/q)\sqrt{(\pi T+i\mu)^2+m^2}} \right\}\right.\nonumber\\
	&+&\left.\frac{2 q}\pi \int_{0}^{\infty }dy\,\frac{h(q,\alpha _{0},y)\tanh(y)}{\cosh (2qy)-\cos(q\pi )}{\cal{R}}\left\{ e^{-2r\cosh(y)\sqrt{(\pi T+i\mu)^2+m^2}} \right\}\right]	\  .
\end{eqnarray}  

In Fig.~\ref{fig1}, we have two plots displaying the thermal FC induced by the cosmic string versus the product $mr$ (left panel) and the ratio $T/m$ (right panel). In both plots we have fixed the ratio $\mu/m=0.5$. Particularly, for the graph in the left panel we have fixed $\alpha_0=0.25$, while for the one in the right we have fixed $mr=1$ and $q=2.5$. Moreover, in the left panel the full curves are given for $q=1.5$, while the dashed ones are given for $q=2.5$. The numbers near these curves represent different values of $T/m$, while the numbers near the curves in the right panel represent different values of $\alpha_0$. As we can see from the plot on the left, $\left\langle \bar{\psi}\psi \right\rangle_{Ts}$ diverges on the string's core for $\sigma>0$ (full curves), while it is finite for $\sigma<0$ (dashed curves). Moreover, this graph shows us that the intensity of the thermal FC induced by the string increases with the temperature. On the other hand, from the plot on the right, we can see that the FC induced by the string vanishes at zero temperature and goes to constant value for large values of $T/m$.
\\

In the discussions above we have assumed that $|\mu |<m$. Now we turn to the case $|\mu |>m$. The FC is an even function of the chemical potential and for definiteness we will assume that $\mu >0$. In this case the calculation of the contribution associated with the antiparticles to the FC  remains the same and the corresponding expression is the half the component   in \eqref{FC_uncomp_2} with exponential $e^{-n\beta\mu}$. So the total thermal FC can be expressed by,
\begin{eqnarray}
	\label{FC_T0}
\left\langle \bar{\psi}\psi \right\rangle_{T}=\left\langle \bar{\psi}\psi \right\rangle_{T}^{(-)}+\frac{mq}{2\pi^2}\sum_j\int_0^\infty dk\int_0^\infty d\lambda \frac{\lambda}{E_\chi} \left[J_{\beta_j}^2(\lambda r)+J_{\beta_j+\epsilon_{j}}^2(\lambda r)\right]\frac1{e^{\beta(E_\chi-\mu)}+1} \ , 
\end{eqnarray}
with $E_\chi=\sqrt{\lambda^2+k^2+m^2}$. In order to analyze the second term on the right hand side of the above equation, we divide the integrals over $k$ and $\lambda$ in two regions corresponding to $\sqrt{k^{2}+\lambda ^{2}}<p_{0}$ and $\sqrt{k^{2}+\lambda^{2}}>p_{0}$ being $p_{0}=\sqrt{\mu ^{2}-m^{2}}$ the Fermi momentum. In the second region we can  use again the expansion (\ref{expansion}). At high temperatures, $T\gg \mu $, the contribution of the states with $E\gg \mu $
dominates and the asymptotic expressions given above are valid. On the other hand in the limit of low temperature the situation is essentially changed. For this case in the limit $T\rightarrow 0$, only the contribution of the states with $E\leqslant \mu $
survive in \eqref{FC_T0}. Defining $\lambda=xp_0$, the integral over $\lambda$ is given in terms of the integral over $x\in [0, \ 1]$ and the integral over $k$ is restricted to the interval $[0, \ p_0\sqrt{1-x^2}]$. The integral over $k$ can be done with use of \cite{Grad}. On basis of these informations,  we get 
\begin{eqnarray}
	\label{FC_T00}
\left\langle \bar{\psi}\psi \right\rangle_{T}\Big|_{T=0}=\frac{mq p_0^2}{2\pi^2}\sum_j\int_0^1  dx x \  {\rm arcsinh} \left(\frac{p_0\sqrt{1-x^2}}{\sqrt{(p_0 x)^2+m^2}}\right) \left(J_{\beta_j}^2(xp_0 r)+J_{\beta_j+\epsilon_{j}}^2(xp_0 r)\right)  \  . \nonumber\\
\end{eqnarray}

Unfortunately we could not find in literature an expression for the integral above. The appearance of a non-vanishing contribution to the FC is related to the presence of particles filling the states with the energies $m\leqslant E\leqslant \mu $. In the absence of the planar angle deficit and magnetic flux, i.e., in the Minkowski spacetime, the summation over $j$ in in the above expression can be explicitly obtained, and we get, 
\begin{equation}
	\label{FC_Min_0}
\left\langle \bar{\psi}\psi \right\rangle_{T}^{(M)}\Big|_{T=0}=\frac{m}	{2\pi^2} \left(\mu p_0-m^2{\rm arctanh} \left({\frac {p_0}{\mu}}\right)\right)  \  .
\end{equation}

\section{Thermal expectation value of the fermionic energy-momentum tensor}
\label{Energy-momentum}
In this section we want to investigate the effects of temperature on the expectation value of the fermionic energy-momentum tensor for the system under consideration. This quantity is important because, in addition to describing the physical structure of the quantum field at a given point, it acts as a source of gravity in the Einstein equations. 

For a charged fermionic field, in the presence of electromagnetic field, the EMT operator
 is expressed as:
\begin{eqnarray}
{\hat{T}}_{\mu \nu }=\frac{i}{2}\left[ \bar{{\hat\psi}}\gamma _{(\mu }{\mathcal{D}}_{\nu
	)}{\hat\psi} -({\mathcal{D}}_{(\mu }\bar{{\hat\psi}})\gamma _{\nu )}\hat{\psi} \right] \ .
\label{EMTdef}
\end{eqnarray}
where ${\mathcal{D}}_{\mu }\bar{\psi}=\partial _{\mu }\bar{\psi}-ieA_{\mu }%
\bar{\psi}-\bar{\psi}\Gamma _{\mu }$ and the brackets in the index
expression mean the symmetrization over the enclosed indices.

The thermal expectation value of the EMT can be evaluated by expanding the field operator, $\hat{\psi}$, as shown in \eqref{psiexp}, and using the relations given in \eqref{traa}. Using both expression, we get
\begin{eqnarray}
	\langle {T}_{\mu\nu}\rangle=\langle {T}_{\mu\nu}\rangle^{(0)}+\langle {T}_{\mu\nu}\rangle^{(+)}+\langle {T}_{\mu\nu}\rangle^{(-)} \ , \label{EMT_Total}
\end{eqnarray}
where the first term in the above expression corresponds to the VEV of the EMT. The latter has been studied in \cite{Mello14a}; so, our main interest resides in the analysis of its thermal corrections coming from particle $(+)$ and antiparticle $(-)$, respectively.  

Similar to the case of the FC, the  thermal corrections can be evaluated by using the mode-sum formula and do not require any renormalization procedure, i.e., all of them are finite. Finally, the thermal corrections are given below:
\begin{eqnarray}
\langle {T}_{\mu\nu}\rangle^{(\pm)}=\pm\frac{i}{2}\sum_\chi\left[ \bar{{\psi}}_\chi^{(\pm)}\gamma _{(\mu }{\mathcal{D}}_{\nu)}{\psi}_\chi^{(\pm)} -({\mathcal{D}}_{(\mu }\bar{{\psi}}_\chi^{(\pm)})\gamma _{\nu )}{\psi}_\chi^{(\pm)} \right]\frac1{e^{\beta(E_\chi\mp\mu)}+1 }  \  . \label{EMT_Thermal}
\end{eqnarray}
In what follow we will evaluate separately all components of this tensor.

\subsection {Energy density}
\label{Energy_density}
Here we first consider the thermal energy density, $\langle T_{0}^0\rangle _T$. For this component  we have, $A_{0}=\Gamma _{0}=0$ and  $\partial_{t}\psi _{\chi }^{(\pm)}=\pm iE_\chi\psi _{\chi }^{(-)}$. Substituting explicitly the expression \eqref{psi+n} for the fermionic modes, and the normalization constant \eqref{C_Nor}, after the
summation over $s$ we obtain
\begin{equation}
	\langle T_{0}^{0}\rangle^{(\pm)} =\frac{q}{(2\pi)^2 }\sum_{j}\int \ dk\int_{0}^{\infty }d\lambda \,\lambda E_\chi[J_{\beta
		_{j}}^{2}(\lambda r)+J_{\beta _{j}+\epsilon _{j}}^{2}(\lambda r)]\frac1{e^{\beta(E_\chi\mp\mu)}+1 }   \ ,
	\label{Thermal_Energy_1}
\end{equation}
with 
\begin{eqnarray}
	E_\chi=\sqrt{m^2+\lambda^2+k^2} \  .
\end{eqnarray}

Let us consider first that $|\mu|<m$. Accepting this case,  we can use the expansion \eqref{expansion} to obtain a more workable expression for total thermal correction defined as, 
\begin{eqnarray}
	\langle T^0_0\rangle_T=\langle T^0_0\rangle^{(+)}+\langle T^0_0\rangle^{(-)} \  .
\end{eqnarray}

After some simple algebraic manipulation, it can be written as,
\begin{eqnarray}
\langle T^0_0\rangle_{T}&=&-\frac{q}{\pi^2 }\sum_{n=1}^\infty(-1)^n\cosh(n\mu\beta)\int_0^\infty d\lambda \lambda \int_0^\infty dk \ e^{-n\beta\sqrt{m^2+\lambda^2+k^2}}\sqrt{m^2+\lambda^2+k^2}\nonumber\\
&\times&\sum_j(J_{\beta	_{j}}^{2}(\lambda r)+J_{\beta _{j}+\epsilon _{j}}^{2}(\lambda r))  \  ,	
\label{Thermal_Energy_2}
\end{eqnarray}
Now to proceed our development, we use the result, 
\begin{eqnarray}
\int_0^\infty dk	e^{-\alpha\sqrt{\rho^2+k^2}}\sqrt{\rho^2+k^2}=\frac{\partial^2}{\partial\alpha^2}\int_0^\infty dk{\frac{{e^{-\alpha\sqrt{\rho^2+k^2}}}} {\sqrt{\rho^2+k^2}}}  \  . \label{Identity}
\end{eqnarray}
Finally using the identity \eqref{identity2}, we get:
\begin{eqnarray}
\langle T^0_0\rangle_T&=&-\frac{q}{\pi^2}\frac1{\sqrt{\pi}}\sum_{n=1}^\infty(-1)^n\cosh(n\mu\beta)\frac{\partial^2}{\partial\alpha^2} \left(\int_0^\infty ds e^{-\alpha^2/(4s^2)-m^2s^2}\right)\Bigg|_{\alpha=n\beta}\nonumber\\
&\times&\int_0^\infty d\lambda \lambda e^{-\lambda^2s^2} \sum_j(J_{\beta	_{j}}^{2}(\lambda r)+J_{\beta _{j}+\epsilon _{j}}^{2}(\lambda r))\int_0^\infty dk \  e^{-k^2s^2} \  . \label{Thermal_Energy_3}
\end{eqnarray}
The integral over the variable $k$ is trivial, and the integral with respect to $\lambda$ can be obtained using \eqref{Int_Bessel_J}. Defining a new variable $z=\frac{r^2}{2s^2}$, the above expressions reads,
\begin{eqnarray}
\langle T^0_0\rangle_T&=&-\frac{q}{4\pi^2r^2}\sum_{n=1}^\infty(-1)^n\cosh(n\mu\beta)\frac{\partial^2}{\partial\alpha^2} \left(\int_0^\infty dz e^{-\alpha^2/(2r^2)z-m^2r^2/(2z)}\right)\Bigg|_{\alpha=n\beta}\nonumber\\
&\times&e^{-z}{\cal{F}}(q.\alpha_0,z) \  .  \label{Thermal_Energy_3}
\end{eqnarray} 
The function ${\cal{F}}(q.\alpha_0,u)$ has explicitly given in \eqref{Sum01} and \eqref{h_func}. Below we give the explicit expression for $e^{-u}{\cal{F}}(q,\alpha_0,u)$:
\begin{eqnarray}
	\label{Expr_1}
	e^{-u}\mathcal{F}(q,\alpha _{0},u) &=&\frac{4}{q}\left[\frac12+\sum_{k=1}^{[q/2]}(-1)^{k}\cos(\pi k/q)\cos (2\pi k\alpha
	_{0})e^{-2u\sin^2(\pi k/q)}\right.\nonumber\\
	&+&\left.\frac q\pi 
	\int_{0}^{\infty }dy\,\frac{h(q,\alpha _{0},y)\sinh y}{\cosh (2qy)-\cos(q\pi )}e^{-2u\cosh^2 (y)} \right] \ .
\end{eqnarray}

Using the expression above, we can see that  the first term on the right hand side provides the thermal energy density in the absence of cosmic string. Taking the integral representation to the Macdonald function, Eq. \eqref{Rep_1}, and the definition \eqref{f-function}, after some intermediate steps we obtain,
\begin{eqnarray}
\langle T^0_0\rangle_T^{(M)}&=&-\frac{2m^4}{\pi^2}\sum_{n=1}^\infty(-1)^n\cosh(n\mu\beta)\left[(m n\beta )^2f_3(m n\beta ) - f_2(m n\beta)\right]  \  . \label{Min_part_0}
\end{eqnarray}
The other terms in the expansion of \eqref{Expr_1}, provide the contributions given by the conical topology of the cosmic string, and the interaction of the charged fermionic field with the magnetic flux running along the string's core. The obtainment of this contribution follows a  similar procedure adopted in the above calculation. The final result is:
\begin{eqnarray}
	\langle T^0_0\rangle_{Ts}&=&-\frac{4m^4}{\pi^2}\sum_{n=1}^\infty(-1)^n\cosh(n\mu\beta) \left\{\sum_{k=1}^{[q/2]}(-1)^{k}\cos(\pi k/q)\cos (2\pi k\alpha
	_{0})\right.\nonumber\\
	&\times&\left[(n\beta m)^2f_3(mu_{kn}) - f_2(mu_{kn})\right]\frac q\pi 
	\int_{0}^{\infty }dy\,\frac{h(q,\alpha _{0},y)\sinh y}{\cosh (2qy)-\cos(q\pi )}\nonumber\\
	&\times&\left[(n\beta m)^2 f_3(mu_{yn})-\left.f_2(mu_{yn})\right]\right\}  \  , \label{Thermal_Energy_4}
\end{eqnarray}
with  the  notations present in \eqref{arguments} and \eqref{index}. The contributions due to the particles and anti-particles in \eqref{Thermal_Energy_4} correspond to the terms with exponents $e^{n\mu\beta}$ and $e^{-n\mu\beta}$, respectively.
	
\subsection{Radial stress}
Our next calculation is the evaluation of the thermal radial stress, $\langle T_{r}^{r}\rangle_T $. For this calculation we take $A_{r}=\Gamma _{r}=0$ in the general definition of the covariant derivative of the fermionic field.
In this way, we have
\begin{equation}
	\langle T_{r}^{r}\rangle_T^{(\pm)} =\mp\frac{i}{2}\sum_{\chi }\left[ \bar{\psi}_{\chi }^{(\pm)}\gamma ^{r}(\partial _{r}\psi _{\chi }^{(\pm)})-(\partial _{r}\bar{\psi}_{\chi }^{(\pm)})\gamma ^{r}\psi _{\chi }^{(\pm)}\right]\frac1{e^{\beta(E_\chi\mp\mu)}+1 } \ .
	\label{Thermal-Radial}
\end{equation}%
Substituting the normalized fermionic wave functions from (\ref{psi+n}) into the above
expression, after some intermediate steps, we arrive at,
\begin{eqnarray}
	\langle T_{r}^{r}\rangle_T &=&\frac{q}{(2\pi)^2}\int_0^\infty d\lambda \ \lambda^3\int  \ dk\frac1{E_\chi}\sum_j{\epsilon_{j}}[J_{\beta _{j}}^{\prime }(\lambda r)J_{\beta_{j}+\epsilon _{j}}(\lambda r)-J_{\beta _{j}}(\lambda r)J_{\beta_{j}+\epsilon _{j}}^{\prime }(\lambda r)]\nonumber\\
	&\times&\left(\frac1{e^{\beta(E_\chi-\mu)}+1 }+\frac1{e^{\beta(E_\chi+\mu)}+1 }\right)\ ,  \label{Thermal-Radial_1}
\end{eqnarray}
where we have developed the summation over the quantum number $s$, and the primes means derivative with respect to the argument of the Bessel functions. The function $S(\lambda r)$ defined below,
\begin{eqnarray}
	S(\lambda r)=\sum_j\epsilon_j\left[J_{\beta _{j}}(\lambda r)J_{\beta_{j}+\epsilon _{j}}^{\prime }(\lambda r)-J_{\beta _{j}}^{\prime }(\lambda r)J_{\beta_{j}+\epsilon _{j}}(\lambda r)\right] \  ,
\end{eqnarray}
can be rewritten as \cite{Mello14a},
\begin{equation}
	S(\lambda r)=\sum_{j}\left[ J_{\beta _{j}}^{2}(\lambda r)+J_{\beta _{j}+\epsilon
		_{j}}^{2}(\lambda r)-\frac{2\beta _{j}+\epsilon _{j}}{\lambda r}J_{\beta _{j}}(x)J_{\beta
		_{j}+\epsilon _{j}}(\lambda r)\right] .  \label{S1}
\end{equation}

In the following calculations we will consider again the case where $|\mu|<m$; so we can write, 
\begin{eqnarray}
\langle T_{r}^{r}\rangle_T &=&-\frac{q}{2\pi^2 }\sum_{n=1}^\infty(-1)^n\cosh(n\mu\beta)\int \ dk\int_0^\infty d\lambda \frac{\lambda^3} {E_\chi}S(\lambda r)e^{-n\beta E_\chi} \  . \label{Thermal-Radial_2}
\end{eqnarray}
Using the identity \eqref{identity2}, we have:
\begin{eqnarray}
\langle T_{r}^{r}\rangle_T &=&-\frac{q}{\pi^2}\sum_{n=1}^\infty(-1)^n\cosh(n\mu\beta)\int_0^\infty \frac{ds}{s} e^{-m^2s^2-(n\beta)^2/(4s^2)}\int_0^\infty d\lambda \lambda^3 S(\lambda r) \  . \label{Thermal-Radial_3}
\end{eqnarray}
Our next calculation is to obtain the integral over the variable $\lambda$ in the equation above. This integral is more elaborate. Now let us start it by direct substitution of \eqref{S1}:
\begin{eqnarray}
	\int_{0}^{\infty }d\lambda \,\lambda ^{3}{e^{-\lambda
			^{2}s^{2}}}S(\lambda r) &=&\sum_{j}\int_{0}^{\infty }d\lambda \,\lambda ^{3}e^{-\lambda
		^{2}s^{2}}\left[ J_{\beta _{j}}^{2}(\lambda r)+J_{\beta _{j}+\epsilon
		_{j}}^{2}(\lambda r)\right]  \notag \\
	&&-\sum_{j}\frac{2\beta _{j}+\epsilon _{j}}{r}\int_{0}^{\infty }d\lambda
	\,\lambda ^{2}e^{-\lambda ^{2}s^{2}}J_{\beta _{j}}(\lambda r)J_{\beta
		_{j}+\epsilon _{j}}(\lambda r)\ .  \label{calc1}
\end{eqnarray}
The first integral can be obtained by taking the derivative $\partial _{s^{2}}$ on the integral  and them use \eqref{Int_Bessel_J}.  As to the second integral we use the formula below,
\begin{equation}
	\int_{0}^{\infty }d\lambda {\lambda ^{2}}{e^{-\lambda ^{2}s^{2}}}\
	J_{\beta _{j}}(\lambda r)J_{\beta _{j}+\epsilon _{j}}(\lambda r)=\frac{	r\epsilon _{j}e^{-y}}{4s^{4}}[I_{\beta _{j}}(y)-I_{\beta _{j}+\epsilon_{j}}(y)]\ ,  \label{Int_Bessel_J_1}
\end{equation}%
with $y=r^{2}/(2s^{2})$. Further, we use the relation
\begin{equation}
	(1+2\epsilon _{j}\beta _{j})[I_{\beta _{j}}(y)-I_{\beta _{j}+\epsilon
		_{j}}(y)]=2\left( y\partial _{y}-y+1/2\right) [I_{\beta _{j}}(y)+I_{\beta
		_{j}+\epsilon _{j}}(y]\ ,  \label{IdentD}
\end{equation}%
to express the term on the right-hand side of \eqref{Int_Bessel_J_1} as the sum of the modified Bessel functions. Combining all these results,  we obtain:
\begin{eqnarray}
	\label{Thermal_Radial_4}
\langle T_{r}^{r}\rangle_T &=&\frac{q}{2\pi^2r^4}\sum_{n=1}^\infty(-1)^n\cosh(n\mu\beta)\int_0^\infty dz z e^{-m^2r^2/(2z)-(n\beta)^22z/(2r^2)}\nonumber\\
&\times& e^{-z}{\cal{F}}(q.\alpha_0,z) \  . 
\end{eqnarray}
Substituting \eqref{Expr_1} to $e^{-z}{\cal{F}}(q.\alpha_0,z)$ into \eqref{Thermal_Radial_4}, we can see that the first term in the right hand side provides the thermal radial stress in Minkowski spacetime. It reads:\footnote{In order to obtain \eqref{Thermal_ Radial_M}, we used the integral representation of Macdonald function, \eqref{Rep_1}, in the integral over the variable $z$.}
\begin{eqnarray}
\langle T_{r}^{r}\rangle_T ^{(M)}=\frac{2m^4}{\pi^2}\sum_{n=1}^\infty(-1)^n\cosh(n\mu\beta)f_2(n\beta m)  \ . \label{Thermal_ Radial_M}
\end{eqnarray}
The following terms provide the contribution induced by the conical topology of the cosmic string and by the magnetic flux. Adopting a similar procedure to obtain the above result, we get:
\begin{eqnarray}
\label{Thermal-Radial_4}
\langle T_{r}^{r}\rangle_{Ts}&=&\frac{4m^4}{\pi^2}\sum_{n=1}^\infty(-1)^n\cosh(n\mu\beta) \left\{\sum_{k=1}^{[q/2]}(-1)^{k}\cos(\pi k/q)\cos (2\pi k\alpha_{0})f_2(m u_{kn})\right.\nonumber\\
&+&\left.\frac q\pi 
\int_{0}^{\infty }dy\,\frac{h(q,\alpha _{0},y)\sinh y}{\cosh (2qy)-\cos(q\pi )} f_2(m u_{yn})\right\}  \ . 
\end{eqnarray}

\subsection{Azimuthal stress}
Our next objective is to calculate the thermal correction to the azimuthal stress. In order to do that we have to observe that $A_{\phi }=q\alpha /e$ and the component of the spin connection reads,
\begin{equation}
	\Gamma _{\phi }=\frac{1-q}{2}\gamma ^{(1)}\gamma ^{(2)}=-\frac{i}{2}(1-q)\Sigma ^{(3)}\ ,\ \Sigma ^{(3)}=\mathrm{diag}(\sigma _{3},\sigma _{3})\
	,  \label{Gam}
\end{equation}%
being $\sigma _{3}$ the Pauli matrix. So this component is given by
\begin{eqnarray}
	\langle T^\phi_\phi\rangle _T^{(\pm)}=\mp\frac i2\sum_{\chi} [\bar{\psi}	_{\chi }^{(\pm)}\gamma ^{\phi }{\mathcal{D}}_{\phi }\psi _{\chi}^{(\pm)}-({%
		\mathcal{D}}_{\phi }\bar{\psi}_{\chi}^{(\pm)})\gamma ^{\phi }\psi _{\chi
	}^{(\pm)}]\frac1{e^{\beta(E_\chi\mp\mu)}+1 }  \  ,  \label{Thermal_Azimu}
\end{eqnarray}
being $\mathcal{D}_{\phi }=\partial_\phi+ieA_\phi+\Gamma_\phi$. In the development of the derivative with respect to the azimuthal variable, we use \eqref{J3} to express $\partial _{\phi }=i\widehat{J}_{3}-i(q/2)\Sigma ^{(3)}$. Moreover, it is easy to show that the anticommutator, $\{\gamma ^{\phi },\Sigma^{(3)}\}$, which appears in the development, vanishes. So, taking in account these results, and substituting the expression for $\gamma^\phi$,  we get after some steps,
\begin{eqnarray}
	\label{Thermal_Azimu_1}
\langle T^\phi_\phi\rangle _T^{(\pm)}&=&-\frac{q^2}{(2\pi)^2r}\sum_{\chi
}\epsilon _{j}(j+\alpha )\frac{\lambda ^{2}}{E_\chi}J_{\beta _{j}}(\lambda
r)J_{\beta _{j}+\epsilon _{j}}(\lambda r)\frac1{e^{\beta(E_\chi\mp\mu)}+1 }  \ .
\end{eqnarray}
Assuming that $|\mu|<m$, the total thermal correction to the azimuthal stress reads,
\begin{eqnarray}
	\langle T^\phi_\phi\rangle_T&=&\frac{2q^2}{\pi^2r}\sum_{n=1}^\infty(-1)^n\cosh(n\mu\beta) \int_0^\infty d\lambda \lambda^2 \int_0^\infty dk \frac{e^{-n\beta\sqrt{m^2+\lambda^2+k^2}}}{\sqrt{m^2+\lambda^2+k^2}}\nonumber\\
	&\times&\sum_j\epsilon_j(j+\alpha ) J_{\beta _{j}}(\lambda
	r)J_{\beta _{j}+\epsilon _{j}}(\lambda r)  \  . \label{Thermal-Azim_cs}
\end{eqnarray}
Using the identity \eqref{identity2}, it is possible to integrate over the $\lambda$ variable by taking \eqref{Int_Bessel_J_1}. The result is,
\begin{eqnarray}
	\label{Thermal-Azim_cs_1}
	\langle T^\phi_\phi\rangle_T&=&\frac{q^2} {\pi^2r^4}\sum_{n=1}^\infty(-1)^n\cosh(n\mu\beta) \int_0^\infty dy y e^{-(n\beta)^2y/(2r^2)-m^2r^2/(2y)-y}\nonumber\\
	&\times&\sum_j(j+\alpha)[I_{\beta _{j}}(y) -I_{\beta _{j}+\epsilon _{j}}(y)] \  , 
\end{eqnarray}
where we have introduced a new variable $y=r^2/(2s^2)$. Observing that $q(j+\alpha )=\epsilon _{j}\beta _{j}+1/2$ and using the relation (\ref{IdentD}), we can write,
\begin{eqnarray}
		\label{Thermal-Azim_cs_2}
	\langle T^\phi_\phi\rangle_T&=&\frac{q^2} {\pi^2r^4}\sum_{n=1}^\infty(-1)^n\cosh(n\mu\beta) \int_0^\infty dy y \ e^{-(n\beta)^2y/(2r^2)-m^2r^2/(2y)}\nonumber\\
	&\times&(y\partial_y+1/2) e^{-y}{\cal{F}}(q,\alpha_0,y) \   .
\end{eqnarray}
Comparing the above result with \eqref{Thermal_Radial_4}, we can verify that the relation below holds,
\begin{eqnarray}
	\label{Rela_1}
\langle T^\phi_\phi\rangle_T=(r\partial_r+1)\langle T^r_r\rangle_T \  .
 \end{eqnarray}

On basis of this relation, we can conclude that the thermal radial stress associated with the Minkowski spacetime, $\langle T^\phi_\phi\rangle^{(M)}_T=\langle T^r_r\rangle^{(M)}_T$, is given by \eqref{Thermal_ Radial_M}. As to the contribution induced by the cosmic string, we get,
\begin{eqnarray}
	\label{Thermal-Azim_cs_3}
\langle {T}^\phi_\phi\rangle_{Ts}&= &\frac{4m^4}{\pi^2}\sum_{n=1}^\infty(-1)^n\cosh(n\mu\beta) \left\{\sum_{k=1}^{[q/2]}(-1)^{k}c_k\cos (2\pi k\alpha_{0})\left[f_2(mu_{kn})\right.\right.\nonumber\\
&-&\left.(2mrs_k)^2f_3(m u_{kn})\right]+\frac q\pi 
\int_{0}^{\infty }dy\,\frac{h(q,\alpha _{0},y)\sinh y}{\cosh (2qy)-\cos(q\pi )}\nonumber\\ &\times&\left.\left[f_2(mu_{yn})-(2mrc_y)^2f_3(mu_{yn})\right]\right\}  \ . 	
\end{eqnarray}

\subsection{Axial stress}

In the calculation of the axial stress, we have to consider $A_{z}=\Gamma_z=0$ in the covariant derivative of the field operator. So we have, ${\mathcal{D}}_{z}\psi _{\chi}^{(\pm)}=\mp ik\psi _{\chi }^{(-)}$. In addition, the matrix $\gamma ^{z}$ coincides with the standard expression for the Dirac matrix in flat spacetime. For this component, we have,
\begin{equation}
	\langle T_{z}^{z}\rangle_T^{(\pm)} =\pm\sum_{\chi }{k}\bar{\psi}_{\chi
	}^{(\pm)}\gamma ^{z}\psi _{\chi }^{(\pm)}\frac1{e^{\beta(E_\chi\mp\mu)}+1}  \ .  \label{Thermal_Axial}
\end{equation}
Substituting the positive/negative fermionic wave-function, \eqref{psi+n}, into the above equation, we can write,
\begin{eqnarray}
	\langle T_{z}^{z}\rangle_T^{(\pm)}=-\frac{q}{8\pi^2}\sum_\chi\frac{\lambda k^2}{E_\chi}(J^2_{\beta_j}(\lambda r)+J^2_{\beta_j+\epsilon_j}(\lambda r)) \frac1{e^{\beta(E_\chi\mp\mu)}+1}  \ .  \label{Thermal_Axial_1}  
\end{eqnarray}

Considering $|\mu|<m$,  the thermal correction to the axial stress reads,
\begin{eqnarray}
\langle T_z^z\rangle_T&=&\frac{q}{\pi^2}\sum_{n=1}^\infty(-1)^n\cosh(n\mu\beta)\int_0^\infty d\lambda \lambda \sum_j (J^2_{\beta_j}(\lambda r)+J^2_{\beta_j+\epsilon_j}(\lambda r))	\nonumber\\
&\times&\int_0^\infty dk k^2\frac{e^{n\beta\sqrt{m^2+\lambda^2+k^2}}}{\sqrt{m^2+\lambda^2+k^2}}  \   . \label{Thermal_Axial_2}  
\end{eqnarray}
Using the identity \eqref{identity2}, it is possible to integrate over $\lambda$. Defining a new variable $y=r^2/(2s^2)$, we obtain after some steps,
\begin{eqnarray}
\langle T_z^z\rangle_T&=&\frac{q}{2\pi^2r^4}\sum_{n=1}^\infty(-1)^n\cosh(n\mu\beta)\int_0^\infty dy y e^{-m^2r^2/(2y)-(n\beta)^2y/(2r^2)}\nonumber\\
&\times&e^{-y}{\cal{F}}(q,\alpha_0,y) \  . \label{Thermal_Axial_3} 
\end{eqnarray}
At this point we note that the above expression has the same structure as the obtained in \eqref{Thermal_Radial_4} for the radial stress. So we conclude
\begin{eqnarray}
	\langle T_z^z\rangle_T=\langle T_r^r\rangle_T  \  . 
\end{eqnarray}

\section{Properties of the energy-momentum tensor and the vacuum energy}

\label{sec5}

In this section we want to investigate the properties obeyed by the thermal expectation value of the EMT. Also the analysis of the thermal energy density only found in previous section in some asymptotic regimes of the physical parameters,  temperature and distance from the string; also we want to discuss its behavior for the case where $|\mu|>m$.

First we can check that the thermal expectation value of the EMT obeys the trace relations below,
\begin{equation}
	\langle T_{\mu }^{\mu }\rangle _{T}=m\langle \bar{\psi}\psi
	\rangle _{T} \  .  \label{TrRel}
\end{equation}
In fact we have shown that the above identity is satisfied for all contributions obtained in previous section.

Another important property obeyed by the thermal expectation value of the energy-momentum is its conservation: $\nabla _{\mu}\langle T_{\nu }^{\mu }\rangle_T =0$. For the problem under consideration this equation is reduced to a single differential equation
\begin{eqnarray}
\langle T_{\phi }^{\phi }\rangle_T=(r\partial_{r}+1)\langle T_{r}^{r}\rangle_T  \  .
\end{eqnarray}
In the development of previous section we have proved that this relation is obeyed by all contributions of the thermal EMT, as shown in \eqref{Rela_1}.

Our next objective in this section is to analyze the behavior of the thermal EMT in some asymptotic regimes of the physical parameters. In order do not prolong this discussion, we will focus only on the energy-density component. 

Let us consider first the behavior of $\langle T^0_0\rangle_{Ts}$ near the string's core. Depending on the relation between the fractional part of the magnetic flux, $\alpha_0$, and the parameter $q$. Two separate analysis  must be considered. For $\sigma<0$, the thermal energy density is finite on the string's core. In this case we can take $r=0$ in 
\eqref{Thermal_Energy_4}. The result is:
\begin{eqnarray}
	\label{energy_asymp_0}
\langle T^0_0\rangle_{Ts}= -\frac{4m^4}{\pi^2}\sum_{n=1}^\infty(-1)^n\cosh(n\mu\beta)[(nm\beta)^2f_3(nm\beta)-f_2(nm\beta)] g(q,\alpha_0) \ ,
\end{eqnarray}
being $g(q,\alpha_0)$ given in \eqref{g_function}.

For $\sigma>0$ the thermal energy density diverges at $r=0$. The divergence comes from the integral part. We can evaluate the behavior of $\langle T^0_0\rangle_{Ts}$ in this region, analyzing the integrand for large values of $y$. Defining a new variable $x=mre^{y}$, the integrand can be written as,
\begin{eqnarray}
	\label{approx0}
	\frac{h(q,\alpha_0,y)\sinh(y)}{\cosh(2qy)-\cos(q\pi)}\approx\frac{\cos[q\pi(1/2-|\alpha_0|)]}{2}\left(\frac x{mr}\right)^\sigma   \  .
\end{eqnarray}
Using the integral representation, \eqref{int_repr},  and after some intermediate steps,\footnote{The procedure developed in this analysis is similar to the one adopted in the thermal condensate.}  we arrive:
\begin{eqnarray}
	\label{energy_asymp_1}
\langle T^0_0\rangle_{Ts}&\approx&-\frac{4qm^4\Gamma(\sigma/2)}{\pi^3}\frac{\cos[q\pi(1/2-|\alpha_0|)]}{(mr)^\sigma} \sum_{n=1}^\infty(-1)^n\cosh(n\mu\beta)(2nm\beta)^{\sigma/2}\nonumber\\
	&\times&[(nm\beta)K_{\sigma/2-3}(nm\beta)-K_{\sigma/2-2}(nm\beta)]  \   .
\end{eqnarray}

In the low temperature regime, $T<m, \ r^{-1}$, two distinct analysis should be take in consideration. For $\sigma<0$, we can  express in the argument of Bessel function, $m\sqrt{(n\beta)^2+(2r\delta)^2}$, with $\delta=s_k, c_y$, as $m\beta\sqrt{n^2+(2rT\delta)^2}$. And now we can take $rT=0$ directly. So we have,
\begin{eqnarray}
	f_\nu(nm\beta)\approx\left(\frac T{nm}\right)^{\nu+1/2}\sqrt\frac\pi2 e^{-m n\beta} \  . 
\end{eqnarray} 
The leading term corresponds to $n=1$. Substituting the approximated expression above into \eqref{Thermal_Energy_4}, we obtain
\begin{eqnarray}
	\label{energy_asymp_2}
\langle T^0_0\rangle_{Ts}&\approx&\left(\frac{2m}{\pi\beta}\right)^{3/2}me^{-(m-|\mu|)\beta}g(q,\alpha_0) \ .
\end{eqnarray}
For $\sigma>0$, we cannot take $r=0$ in the argument of the Bessel function. In fact in this limit the energy-density diverges, and the divergence comes from the integral contribution in \eqref{Thermal_Energy_4}.  Here we have to take an approximated expression for the integrand considering large values of $y$. Defining a new variable $x=rTe^y$, an approximated expression for the term inside the integral is obtained:
\begin{eqnarray}
	\label{approx1}
	\frac{h(q,\alpha_0,y)\sinh(y)}{\cosh(2qy)-\cos(q\pi)}\approx\frac{\cos[q\pi(1/2-|\alpha_0|)]}{2}\left(\frac x{rT}\right)^\sigma   \  .
\end{eqnarray}
As to the  functions $f_3$ and $f_2$, defined in \eqref{f-function},  we use the integral representation \eqref{Rep_1} for the modified Bessel function. So we can write,
\begin{eqnarray}
	f_\nu(m\sqrt{(n\beta)^2+(2rc_y)^2})\approx\frac {2^{\nu-1}}{(m\beta)^{2\nu}}\int_0^\infty dz z^{\nu-1} e^{-(m\beta)^2/(4z)-(n^2+x^2)z}  \  .
\end{eqnarray} 
Adopting the above representation in the integral of the energy-density, and carrying out some other intermediate steps, we arrive that the leading contribution is given by
\begin{eqnarray}
\label{energy_asymp_3}
\langle T^0_0\rangle_{Ts} &\approx&\frac{m^4q\Gamma(\sigma/2)2^{\sigma/2-2}}{\pi^3} \sqrt{\frac{\pi}{2}}\frac{\cos[q\pi(1/2-|\alpha_0|)]}{(rT)^\sigma} \left(\frac Tm\right)^{(3+\sigma)/2}e^{-(m-|\mu|)/T} \  .
\end{eqnarray}

As a final remark regarding the low temperature limit, it worths to note that the energy density \eqref{energy_asymp_2} and the corresponding result found for a scalar field in the same setup \cite{Oliveira23} have the same dependency with the temperature (for $D=3$ in the scalar field case).

At high temperature limit, $T\gg m, \ r^{-1}$, the main contribution in the energy-density comes from the large values of $n$. Unfortunately our previous expression is not convenient to develop this investigation. In order to evaluate this behavior we use a similar procedure as we did in the study of FC. First we take the expression  \eqref{Resum} with $b= \{ 2r\sin(\pi k/q), \ 2r\cosh(y)\}$, $w_{n}=\sqrt{(2\pi n+\alpha)^2/\beta^2+m^2}$ with $\alpha=\pi+ i\beta{\mu}$. The obtainment of the expressions  requires many intermediate steps. In order do not make this discussion here too long, in the Appendix \ref{energy_high_T} we give the most relevant details about the procedure developed. Below we present only our final result:
\begin{eqnarray}
\label{energy-asymp_4}
\langle T^0_0\rangle_{Ts} &\approx&\frac{2\pi T^3}{r}\left\{\sum_{k=1}^{[q/2]}(-1)^{k}\cot(\pi k/q)\cos (2\pi k\alpha
_{0})e^{-2\pi r s_kT}\cos(2r\mu s_k)\right.\nonumber\\
&+&\left.\frac q\pi  \int_{0}^{\infty }dy\,\frac{h(q,\alpha _{0},y)\tan (y)}{\cosh (2qy)-\cos(q\pi )}e^{-2\pi r c_yT}\cos(2r\mu c_y)\right\} \  .
\end{eqnarray}
From this expression we can see that it does not depend on the mass of the field, and that it decays exponentially with $T$. Moreover, for fixed $T$, the energy-density goes to zero for points far from the string. In addition, note that this fermionic energy density decays as the temperature increases and is in contrast with the result found for a scalar field in the same setup \cite{Oliveira23}, which increases linearly with the temperature.

Finally we want to say that this result is in agreement with the corresponding one found by Linet in \cite{Linet96}. Considering the Minkowskian thermal contribution to the energy-density given in \eqref{Min_part_0}, in the limit that $m=\mu=0$, we  have 
\begin{eqnarray}
\langle T^0_0\rangle_T^{(M)}=\frac{7\pi^2T^4}{60} \  .
\end{eqnarray}
So we conclude that the dominant thermal contribution to the fermionic energy-density is given by the Minkowskian part.

\begin{figure}[!htb]
	\begin{center}
		\centering
		\includegraphics[scale=0.4]{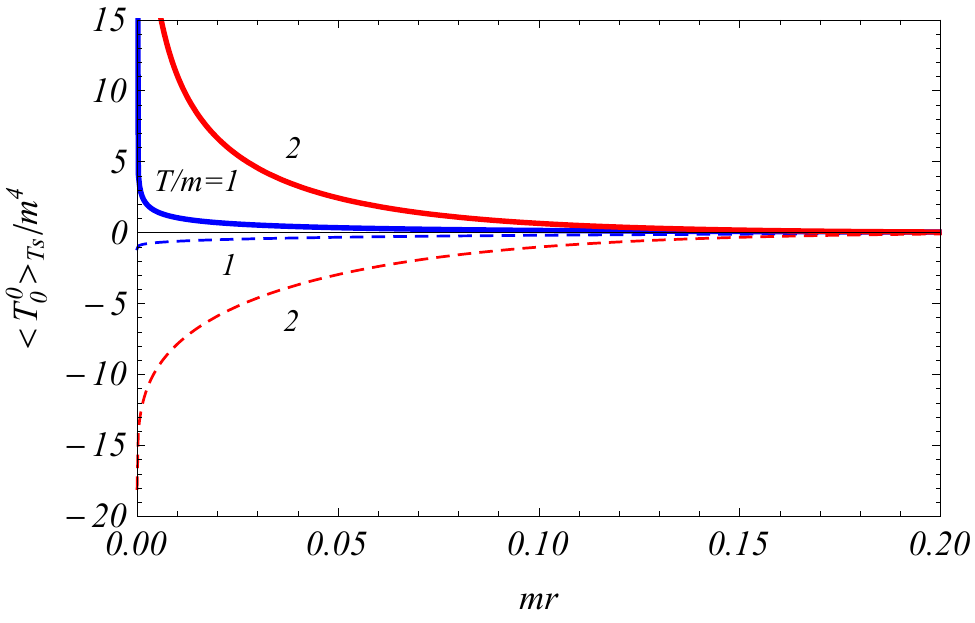}
		\quad
		\includegraphics[scale=0.4]{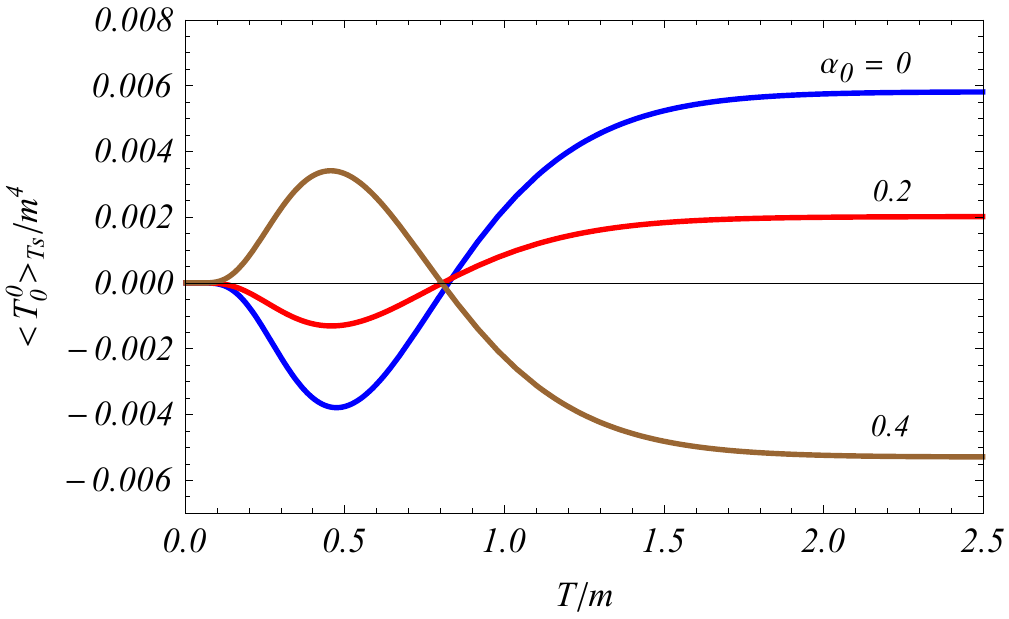}
		\caption{The thermal expectation value of energy density induced by the string is plotted as function of the ratios $mr$ (left panel) and $T/m$ (right panel). $\alpha_{0}=0.25$ is fixed for the plot in the left panel, while $rm=1$ and $q=2.5$ are fixed for the right one. In both plots $\mu/m=0.5$ is fixed. In the left panel, the full curves are given for $q=1.5$ and the dashed ones, for $q=2.5$. The numbers near these curves represent different values of the ratio $T/m$, while the numbers near the curves in the right panel represent different values of $\alpha_0$.}
		\label{fig3}
	\end{center}
\end{figure}

In the Fig.~\ref{fig3}, the thermal expectation value of the energy density induced by the string is plotted for product $mr=1$ (left panel) and the ration $T/m$ (right panel). For the plot in the left panel we have fixed $\alpha_{0}=0.25$, while $mr=1$ and $q=2.5$ for the one in the right panel. Moreover, in the left panel the full curves are plotted for $q=1.5$ and the dashed ones for $q=2.5$. The numbers near the curves in both plots represent different values of $T/m$ and $\alpha_{0}$ in the left and right panels, respectively. Starting from the left panel, we note that the thermal energy density induced by the string is divergent at the string's core for $q=1.5$ corresponding to $\sigma>0$ (full curves) while it is finite for $q=2.5$ corresponding to $\sigma<0$ (dashed curves). Another detail of this plot is the intensity of the energy density increasing with the temperature. As to the right panel, we can see that the $\langle T^0_0\rangle_{Ts}$ is zero at $T/m=0$ and goes to a constant value for large values of temperature.
\\ 

In the discussion above we have assumed that $|\mu |<m$. Now we turn to the case $|\mu |>m$. The VEV of the EMT is an even function of the chemical potential. So for definiteness we will assume that $\mu >0$. In this case the calculation of the contribution associated with the antiparticles to the energy-density remains the same and the corresponding expression is given the contribution in \eqref{Thermal_Energy_3} with negative exponential factor $e^{-n\mu\beta}$, $\langle T^0_0\rangle_{T}^{(-)}$. So the total  energy-density is expressed by,
\begin{eqnarray}
	\label{energy_density_high_T}
\langle T^0_0\rangle_{T}=\langle T^0_0\rangle_T^{(-)}+\frac{2q}{(2\pi)^2}\int_0^\infty dk\int_{0}^{\infty }d\lambda \,\lambda E_\chi\sum_{j}[J_{\beta
	_{j}}^{2}(\lambda r)+J_{\beta _{j}+\epsilon _{j}}^{2}(\lambda r)]\frac1{e^{\beta(E_\chi-\mu)}+1 }\  ,
\end{eqnarray}
with $E_\chi=\sqrt{m^2+k^2+\lambda^2}$. 

The integral over $k$ and $\lambda$ are divided in two regions: $\sqrt{\lambda^2+k^2}<p_0$ and $\sqrt{\lambda^2+k^2}>p_0$, being $p_0=\sqrt{\mu^2-m^2}$. For the second region we can use the expansion $\frac1{e^y+1}=-\sum_{n=1}^\infty(-1)^ne^{-ny}$. For high temperature, $T\gg\mu$, the states with $E>\mu$ dominate, and the asymptotic expression given before is still valid; however for low temperature, $T\to 0$, only states with $E<\mu$ survive. Defining $\lambda= xp_0$, we change the integral over $\lambda$ by the integral over $x\in[0, \ 1]$. As to the integral over $k$ it will be carried out in the interval $[0, \ p_0\sqrt{1-x^2}]$. Taking into account these restrictions, we can evaluate the integral over $k$, resulting in
\begin{eqnarray}
	\label{energy_T0}
\langle T^0_0\rangle_{T}\Big|_{T=0}&=&\frac{qp_0^2}{(2\pi)^2}\int_0^1 dx x \left[(m^2+p_0^2x^2)\ln\left(\frac{\mu+p_0\sqrt{1-x^2}}{\sqrt{(p_0x)^2+m^2}}\right)\right.\nonumber\\
&+&\left.p_0\mu\sqrt{1-x^2}\right] \sum_{j}[J_{\beta
	_{j}}^{2}(xp_0 r)+J_{\beta _{j}+\epsilon _{j}}^{2}(xp_0 r)]\  . 
\end{eqnarray}

Unfortunately we did not find in literature the expression for the integral above. The appearance of a non-vanishing result for the thermal energy-density at $T=0$, is related to the presence of particle filling states with energies $m\leq E\leq\mu$. In the absence of the planar angle deficit and magnetic flux, i.e., in the Minkowski spacetime, the summation over $j$ in in the above expression can be explicitly obtained, and after some intermediate steps, we get: 

\begin{eqnarray}
\langle T^0_0\rangle_{T}^{(M)}\Big|_{T=0}&=&-\frac1{8{(2\pi)}^{2}}\left[{m}^{4}\ln  \left( {
	\frac {{m}^{2}+{p_{{0}}}^{2}}{{m}^{2}}} \right)+2 \left( 2{m}^{2}
{\mu}^{2}-{\mu}^{4}+2{\mu}^{2}{p_{{0}}}^{2} \right) \ln  \left( {
\frac {\mu+p_{{0}}}{\mu}} \right) \right.\nonumber\\
&+&p_0^2\left( 2{m}^{2}+{p_{{0}}}^{2} \right) \ln  \left( {\frac {{m}^{2}+{p_{{0}}}^{2}}{ \left(\mu+p_{{0}} \right) ^{2}}}\right)+{p_{{0}}}^{2} \left( {m}^{2}-{\mu}
	^{2}+{p_{{0}}}^{2} \right)\nonumber\\
	&+&\left.2\mu p_{{0}} \left(-2{m}^{2}+{\mu}^{
2}-3{p_{{0}}}^{2} \right) \right] \  . 
\end{eqnarray}

\begin{figure}[!htb]
	\begin{center}
		\centering
		\includegraphics[scale=0.4]{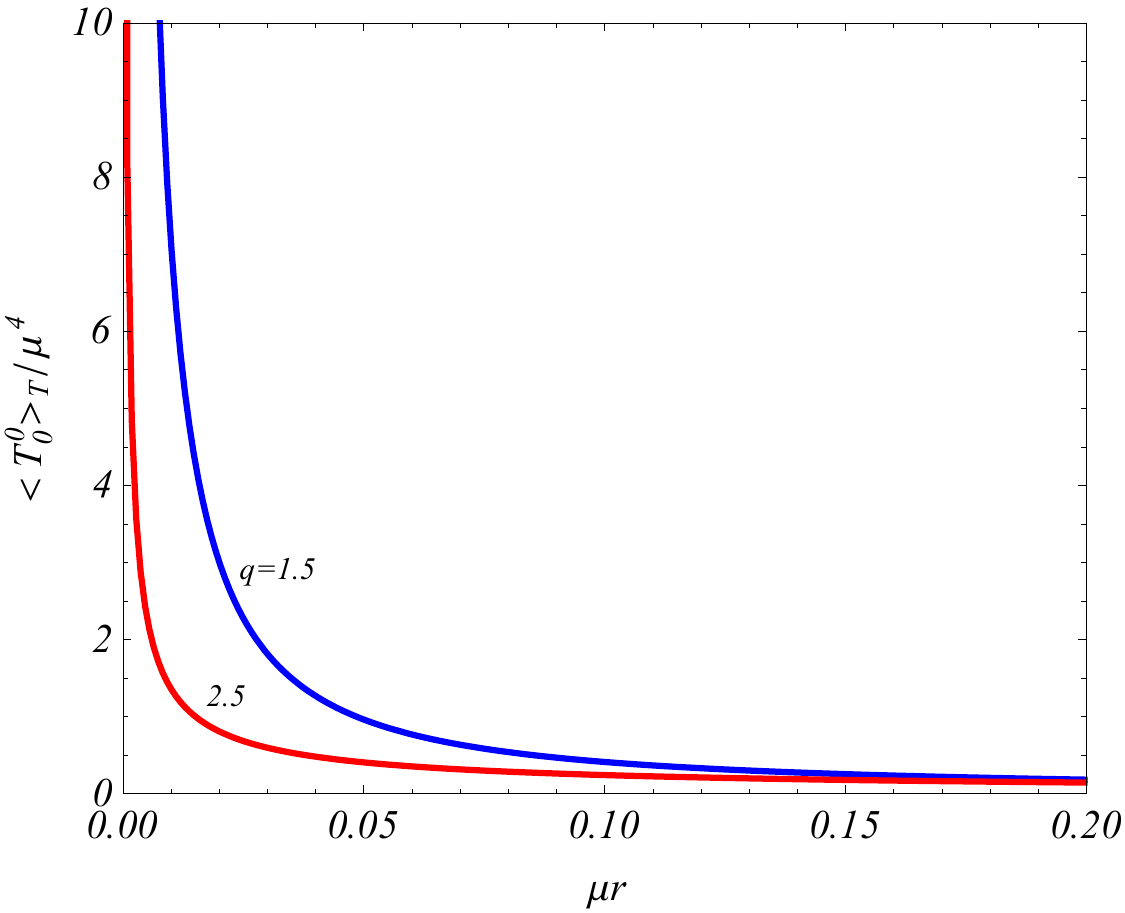}
		\caption{The total energy density in \eqref{energy_T0} for the case $\mu>m$ is plotted as function of $\mu r$. $\alpha_{0}=0.25$ is fixed and the numbers near the curves represent different values of $q$.}
		\label{fig4}
	\end{center}
\end{figure}

In Fig.~\ref{fig4}, we have plotted the total energy density Eq.~\eqref{energy_T0} for a massless scalar field as function of the product $\mu r$, with fixed $\alpha_{0}=0.25$. The numbers near the curves correspond to different values of the parameter associated to the deficit angle, $q$.

\section{Conclusions}
\label{conc} 
In this paper, we have considered the combined effects of the finite temperature, electromagnetic interaction and conical topology on the analysis of the FC and the EMT associated with a charged fermionic field in thermal equilibrium at finite temperature, $T$. We have assumed that the conical topology of the spacetime is due to the presence of a cosmic string. Moreover, we have considered non-vanishing fermionic chemical potential, $\mu$. In the analysis two distinct situations have been considered: the modulus of the chemical potential smaller than the fermionic mass, $|\mu|<m$, and  greater than it, $|\mu|>m$. For the first case, the analysis were more elaborate, and we were able to provide closed expressions for the FC and EMT. As to the second situation, considering a positive chemical potential, we explicitly have found non-vanishing values for the FC and energy-density in the limit $T\to 0$. The appearance of these observables is related to the presence of particles filling the states with the energies $m\leqslant E\leqslant \mu $. 

The analysis of the thermal fermionic condensate has been developed in Section \ref{FCondensate}, by taking the ensemble average of the operator ${\bar{\psi}\psi}$ expressed in terms of a complete set of fermionic modes. This average takes into account contributions of particles and anti-particles. Considering first the case with $m>|\mu|$, we expanded the thermal factor, according to \eqref{expansion}.  This term was given by a sum of the contribution in pure Minkowski spacetime, \eqref{FC_Min}, plus a contribution induced by the presence of the cosmic string, \eqref{FC_cs1}. Both quantities are even functions of the chemical potential.  The contribution induced by the cosmic string has been analyzed in various limiting regions. Near the string and considering $\sigma=(2|\alpha_0|-1)q+1<0$, the result was easily obtained in \eqref{FC_cs2}; for $\sigma>0$ the FC diverges near the string and its behavior is given in \eqref{FC_cs3}. At low temperature and considering $\sigma<0$, we obtained \eqref{FC_cs4}. For $\sigma>0$ the behavior is given by \eqref{FC_cs5}. At this point we would like to mention that both behaviors decay exponentially as $e^{-(m-|\mu|)/T}$. The behavior of the thermal FC in the high temperature regime is given in \eqref{high_Temp}. In this regime the FC decays exponentially with $T$. Finally we have analyzed the FC for the case where $|\mu|>m$. Specifically we consider the case of positive chemical potential. We found that the contribution associated with antiparticle remains the same; however the contribution due to the particles is more delicate. We had to separate the integral over the variables $k$ and $\lambda$ in \eqref{FC_T0} in two regions. Considering the limit $T\to0$, the corresponding expression is given in \eqref{FC_T00}. 

The analysis of thermal expectation value of the operator EMT has been presented in Section \ref{Energy-momentum}. We have explicitly exhibited in detail the calculations of all non-vanishing components, considering $|\mu|<m$.
The most important properties obeyed by the thermal expectation value of the EMT have been proved in Section
\ref{sec5}. There we have presented the thermal energy-density for several asymptotic regimes of the parameters of the system. For points near the string we found \eqref{energy_asymp_0} and \eqref{energy_asymp_1} for $\sigma<0$ and $\sigma>0$, respectively, where the latter diverges as $1/(mr)^\sigma$.  In the low temperature and considering $\sigma<0$ the energy-density behaves as \eqref{energy_asymp_2}, and for $\sigma>0$ as \eqref{energy_asymp_3}. We can observe that both behaviors present an exponential decay. At high temperature the energy-density behaves as \eqref{energy-asymp_4}. This expression does not depend on the mass of the fermion and decays with exponentially with $T$. Finally we have analyzed the thermal energy-density considering positive chemical potential with $\mu>m$. For this case, in the limit of $T\to0$ the contribution of the particle survives and the corresponding result is given in \eqref{energy_T0}. 

As general information about our results, we see that all the thermal physical observable are even functions of the fractional part of the ratio of the magnetic flux by the quantum one, $\alpha_0$, even the expressions for the FC and energy-density given by \eqref{high_Temp} and \eqref{energy-asymp_4}. This is typical Aharonov-Bohm effect. This fact occur because, in our system, we have considered the cosmic string without inner structure. In \cite{Mikael_17}, the analysis of the VEV of fermionic energy-momentum tensor at zero temperature  have been obtained considering a finite extension for the magnetic flux. There it was shown that, for points outside the core, the corresponding VEV is given by a contribution associated to an idealized cosmic string, plus another contribution induced by the core. The latter depend on the total magnetic flux. So it is expected that at nonzero temperature the thermal corrections also present a core-induced contribution. It is our objective in the near future to develop this analysis.
	
Although all the analytical results obtained in this paper, have be developed considering  $|\mu|<m$, in high-temperature regime, the expressions obtained for FC and energy-density remain valid for $|\mu|>m$, as long as $T>|\mu|$, as explained in the text below Eq.s \eqref{FC_T0} and \eqref{energy_density_high_T}. So, we understand that under extreme conditions, i.e., for $|\mu|>T$, the equations \eqref{high_Temp} and \eqref{energy-asymp_4} are not adequate to represent the desired behavior.

A more subtle dependence in our results is related with the parameter $\sigma=(2|\alpha_0|-1)q+1$. As we could notice the asymptotic behavior of FC and energy-density depend crucial on this parameter. Being $\alpha_0=0$, we have $\sigma<0$. For this case there are no explicit divergences in the FC or energy-density for points near the string or low temperature. In the absence of conical topology, i.e., $q=1$, $\sigma>0$, the above mentioned asymptotic limits present divergences.

\section*{Acknowledgments}
W.O.S is supported under grant 2022/2008, Paraíba State Research Foundation (FAPESQ). E.R.B.M is partially supported by CNPq under Grant no 301.783/2019-3.

\appendix
\section{Details of the calculation of the $\langle \bar{\psi}\psi\rangle_{Ts}$ near the string in the case $\sigma>0$}
\label{FC_r0_appendix}
In this appendix we give some details on calculation of the FC near the string for the divergent case, $\sigma>0$, given by \eqref{FC_cs3}. Other analysis in this paper follows the same procedure.

This divergent behavior comes from the integral over $y$ in \eqref{FC_cs1}. Thus, taking the asymptotic expression \eqref{asymp_y} and the argument $m\sqrt{(n\beta)^2+(re^y)^2}\approx\sqrt{(nm\beta)^2+(mre^y)^2}$, we can make the change of variable $x=mre^y$, getting the following integral:
\begin{equation}
	I_1=\frac{\cos[(1/2-|\alpha_{0}|)q\pi]}{2(mr)^\sigma }\int_{mr}^{\infty}dxx^{\sigma-1}f_{1}(\sqrt{(nm\beta)^2+x^2}) \ .
	\label{A1}
\end{equation}

Now using the integral representation for the Macdonald function given in \eqref{int_repr}, we get
\begin{equation}
	I_1=\frac{\cos[(1/2-|\alpha_{0}|)q\pi]}{2(mr)^\sigma}\int_{0}^{\infty}dze^{-1/(4z)-(nm\beta)^2z}\int_{mr}^{\infty}dxx^{\sigma-1-x^2z} \ .
	\label{A2}
\end{equation}
Since we are interested in the behavior near the string's core, its safe to take $mr\approx0$ in the lower limit of integration over $x$. This allow us to integrate over $x$, providing us the following result:
\begin{equation}
	I_1=\frac{\cos[(1/2-|\alpha_{0}|)q\pi]}{4(mr)^\sigma}\Gamma\left(\frac{\sigma}{2}\right) \int_{0}^{\infty}\frac{dz}{z^{\sigma/2}}e^{-1/(4z)-(nm\beta)^2z} \ .
	\label{A3}
\end{equation}
We note that integration over $z$ in the expression above can be performed with the use of integral representation \eqref{Rep_1}, by identifying $d=-\sigma/2$, $a=1/2$ and $b=nm\beta$. This allow us to rewrite \eqref{A3} as:
\begin{equation}
	I_1=2^{\sigma/2-2}\frac{\cos[(1/2-|\alpha_{0}|)q\pi]}{4(mr)^\sigma}\Gamma(\sigma/2)\frac{K_{\sigma/2-1}(mn\beta)}{(mn\beta)^{\sigma/2-1}} \ ,
\end{equation}
which plugged in \eqref{FC_cs1} gives us the result found in \eqref{FC_cs3}.
\section{Details of the calculation of the $\langle T^0_0\rangle_{Ts}$ in the limit of high temperature}
\label{energy_high_T}
To provide the behavior of $\langle T^0_0\rangle_{Ts}$ in the limit of high temperature, we need to develop two distinct terms in \eqref{Thermal_Energy_4} separately. The first one is
\begin{eqnarray}
	\label{res_1}
	\sum_{n=1}^\infty(-1)^n\cosh(n\mu m)(mn\beta)^2f_3(m\sqrt{(n\beta)^2+b^2})=m^2\frac{d^2}{d\mu^2}\sum_{n=1}^\infty \cos(\alpha n)f_3(m\sqrt{(n\beta)^2+b^2}) \  ,
\end{eqnarray} 
with  $b= \{ 2r\sin(\pi k/q), \ 2r\cosh(y)\}$ and $\alpha=\pi+ i\beta{\mu}$.

Defining by $S_1$ the summation on the right-hand side of the above expression, the leading term of its asymptotic behavior is 
\begin{eqnarray}
	S_1\approx\frac{\sqrt{2\pi}}{\beta m^6}{\cal{R}}\{[(\pi/\beta+i\mu)^2+m^2]^{5/2}f_{5/2}(b\sqrt{(\pi/\beta+i\mu)^2+m^2})\}  \  .
\end{eqnarray}
Considering now $T\gg m, \ \mu$, and using the approximated expression for the modified Bessel function for large arguments \cite{Abra}, we have, 
\begin{eqnarray}
	S_1\approx \frac{(\pi T)^3}{m^6b^3}e^{-b\pi T}\cos(b\mu) \  .
\end{eqnarray}
Consequently we get,
\begin{eqnarray}
	\label{cont_1}
	m^2\frac{d^2S_1}{d\mu^2}\approx-\frac{(\pi T)^3}{m^4b}e^{-b\pi T}\cos(b\mu)  \  . 
\end{eqnarray}

The second contribution 
\begin{eqnarray}
	\label{res_2}
	S_2=\sum_{n=1}^\infty(-1)^n\cosh(n\mu m)f_2(m\sqrt{(n\beta)^2+b^2})
\end{eqnarray}
behaves as,
\begin{eqnarray}
S_2=\frac{\sqrt{2\pi}}{\beta m^4}{\cal{R}}\{w_0e^{-bw_0}\}   \  ,
\end{eqnarray}
with $w_0=\sqrt{(\pi/\beta+i\mu)^2+m^2}$. Considering $T\gg\mu$, we have,
\begin{eqnarray}
	S_2\approx\frac{(\pi T)^2}{m^4b^2}e^{-b\pi T}\cos(b\mu) \  .
\end{eqnarray}

The sum of both contributions is:
\begin{eqnarray}
		m^2\frac{d^2S_1}{d\mu^2}-S_2\approx-\frac{(\pi T)^2}{m^4b}e^{-b\pi T}\cos(b\mu)\left(\pi T+\frac 1 b\right)
\end{eqnarray}

Because we are working in the limit of $T\gg r^{-1}$, the first term inside the parenthesis in the r.h.s. of the above expression is the leading contribution.


\begin{thebibliography}{99}

\bibitem{Kibble} T. W. Kibble, J. Phys. A. {\bf 9}, 1387 (1976).

\bibitem{V-S} A. Vilenkin and E. P. S. Shellard, {\it Cosmic Strings and Other Topological Defects} (Cambridge University Press, Cambridge, England, 1994).

\bibitem{scalar} B. Linet, Phys. Rev. D \textbf{35}, 536 (1987).

\bibitem{scalar1} A. G. Smith, in \textit{Symposium on the Formation and
	Evolution of Cosmic String}, edited by G. W. Gibbons, S. W. Hawking and T.
Vachaspati (Cambridge University Press, Cambridge, England, 1989).

\bibitem{scalar2} P. C. Davies and V. Sahni, Class. Quantum Grav. \textbf{5}%
, 1 (1987).

\bibitem{scalar3} T. Souradeep and V. Sahni, Phys. Rev. D \textbf{46}, 1616
(1992).

\bibitem{scalar4} M. E. X. Guimar\~{a}es and B. Linet, Class. Quantum Grav.
\textbf{10}, 1665 (1993).

\bibitem{Aram1} E. R. Bezerra de Mello, V. B. Bezerra, A. A. Saharian and A.
S. Torloyan, Phys. Rev. D \textbf{74}, 025017 (2006).

\bibitem{Site11} Yu. A. Sitenko and N. D. Vlasii,     Class. Quant. Grav. {\bf 29},   095002 (2012).

\bibitem{ferm} V. P. Frolov and E. M. Serebriany, Phys. Rev. D \textbf{15},
3779 (1987).

\bibitem{ferm1} B. Linet, J. Math. Phys. \textbf{36}, 3694 (1995).

\bibitem{ferm3} V. B. Bezerra and N. R. Khusnutdinov, Class. Quantum Grav.
\textbf{23}, 3449 (2006).

\bibitem{Beze08f} E. R. Bezerra de Mello, V. B. Bezerra, A. A. Saharian and A. S. Tarloyan, Phys. Rev. D \textbf{78}, 105007 (2008).

\bibitem{Skarzhinsky} V. D. Skarzhinsky, D. D. Harari and U. Jasper, Phys. Rev. D \textbf{49}, 755 (1994).

%\bibitem{Cast09} A.H. Castro Neto, F. Guinea, N.M.R. Peres, K.S. Novoselov, and A.K. Geim, Rev.Mod. Phys. \textbf{81}, 109 (2009).

\bibitem{Braganca_15}  E. A. F. Bragan\c{c}a, H. F. Santana Mota, E. R. Bezerra de Mello, Int. J. Mod. Phys. D \textbf{24}, 1550055 (2015).

\bibitem{Braganca_19} E. A. F. Bragan\c{c}a, H. F. Santana Motta and E. R. Bezerra de Mello, Eur. Phys. J. Plus. {\bf 134}, 400 (2019).

\bibitem{Mello13a}E. R. Bezerra de Mello and A. A. Saharian,  Eur. Phys. J. C {\bf 73}, 2532 (2013). 


\bibitem{Mello14a} A. Bellucci, E. R. Bezerra de Mello and A. A. Saharian,   Eur. Phys. J. C {\bf 74},  2688 (2014).

\bibitem{Mohammadi_16} A. Mohammadi and E. R. Bezerra de Mello, Phys. Rev. D {\bf 93}, 123521 (2016).

\bibitem{Oliveira23} W. Oliveira dos Santos and E. R. Bezerra de Mello, Eur. Phys. J. C. {\bf 83}, 163 (2023).

\bibitem{Linet96} B. Linet, Class. Quantum Grav. {\bf 13}, 97 (1996).

\bibitem{Braganca_16} S. Bellucci, E. R. Bezerra de Mello, E. Bragan\c{c}a and A. A. Saharian, Eur. Phys. J. C {\bf 76}, 350 (2016). 

\bibitem{Azadeh_15} A. Mohammadi, E. R. Bezerra de Mello and A. A. Saharian,     J. Phys. A {\bf 48}, 185401 (2015).

\bibitem{Abra} M. Abramowitz and I. A. Stegun, \textit{Handbook of Mathematical Functions} (Dover, New York, 1972).

\bibitem{BezerradeMello:2010ii}
E.~R.~Bezerra de Mello, V.~B.~Bezerra, A.~A.~Saharian and V.~M.~Bardeghyan,
%``Fermionic current densities induced by magnetic flux in a conical space with a circular boundary,''
Phys. Rev. D \textbf{82}, 085033 (2010).

\bibitem{Bell14T} S. Bellucci, E.R. Bezerra de Mello and A.A. Saharian, Phys. Rev. D \textbf{89}, 085002 (2014).

%\bibitem{Saha_2010} S. Bellucci, A. Saharian, and V. Bardeghyan, Phys. Rev. D {\bf 82},  065011 (2010).

%\bibitem{SahaRev} A.A. Saharian, \textit{The Generalized Abel-Plana Formula with Applications to Bessel Functions and Casimir Effect} (Yerevan State University Publishing House, Yerevan, 2008); Preprint ICTP/2007/082; arXiv:0708.1187.

\bibitem{Grad} I. S. Gradshteyn and I. M. Ryzhik. \textit{Table of Integrals, Series and Products}
(Academic Press, New York, 1980).

\bibitem{Beze10b} E. R. Bezerra de Mello, V. B. Bezerra, A. A. Saharian, and V. M. Bardeghyan, Phys. Rev. D \textbf{82}, 085033 (2010).

\bibitem{Bellucci:2009jr} S. Bellucci and A. A. Saharian, Phys. Rev. D \textbf{79}, 085019 (2009).

\bibitem{Cruz:2018bqt} M. B. Cruz, E. R. Bezerra de Mello and A. Y. Petrov, Mod. Phys. Lett. A \textbf{33}, no.20, 1850115 (2018).

\bibitem{Mikael_17} M.~S.~Maior de Sousa, R.~F.~Ribeiro and E.~R.~Bezerra de Mello, Phys. Rev. D \textbf{95}, no.4, 045005 (2017).



\end{thebibliography}
\end{document}